\documentclass[journal]{IEEEtran}
\usepackage{cite}
\usepackage{amsmath,amssymb,amsfonts,amsthm}

\usepackage[utf8]{inputenc} 
\usepackage[T1]{fontenc}

\usepackage{graphicx}
\usepackage{textcomp}
\usepackage{xcolor,soul}
\usepackage{stmaryrd}
\usepackage{algpseudocode}
\usepackage{amsmath}
\usepackage{mathrsfs}
\usepackage{bm}
\usepackage[abs]{overpic}
\usepackage[nocomma]{optidef}

\usepackage{physics}
\usepackage{tikz}
\usepackage{mathdots}
\usepackage{yhmath}
\usepackage{cancel}
\usepackage{siunitx}
\usepackage{array}
\usepackage{multirow}
\usepackage{amssymb}
\usepackage{gensymb}
\usepackage{tabularx}
\usepackage{extarrows}
\usepackage{booktabs}
\usepackage[ruled,linesnumbered]{algorithm2e}
\makeatletter
\newcommand{\mathleft}{\@fleqntrue\@mathmargin0pt}
\newcommand{\mathcenter}{\@fleqnfalse}
\makeatother

\usetikzlibrary{fadings}
\usetikzlibrary{patterns}
\usetikzlibrary{shadows.blur}
\usetikzlibrary{shapes}

\theoremstyle{plain}
\newtheorem*{theorem*}{Theorem}

\newtheorem{remark}{Remark}
\newtheorem*{eg}{Example}

\usepackage{xcolor,varwidth}

\usepackage[scaled=.75]{beramono}
\newcommand{\bpara}[1]		{\medskip \noindent {\bf #1}}

\renewcommand\geq\geqslant
\renewcommand\leq\leqslant

\newcommand\fig[1]			{Fig.~\ref{#1}}

\def\eg					    {\emph{e.g.,}~}
\def\ie					    {\emph{i.e.,}~}

\usepackage{xspace}

\DeclareDocumentCommand{\sd}{o}  
{{\underline{\ast}\IfValueT{#1}{_{#1}}}}

\DeclareDocumentCommand{\xmod}{m o o o}  
{%
\IfNoValueTF{#4}
{{#1}\IfValueT{#2}{_{m,\mathsf{#2}}}\IfValueT{#3}{#3}}
{{#1}\IfValueT{#2}{_{m,\mathsf{#2}}^{\mathsf{#4}}}\IfValueT{#3}{#3}}
}

\usepackage[inline]{enumitem}
\DeclareMathAlphabet{\mathsfit}{T1}{\sfdefault}{\mddefault}{\sldefault}
\SetMathAlphabet{\mathsfit}{bold}{T1}{\sfdefault}{\bfdefault}{\sldefault}
\usepackage[switch]{lineno}
\usepackage{siunitx}
\def\BibTeX{{\mathrm B\kern-.05em{\sc i\kern-.025em b}\kern-.08em
    T\kern-.1667em\lower.7ex\hbox{E}\kern-.125emX}}
    
\usepackage{hyperref}
\usepackage{etoolbox}

\begin{document}

%\title{Max-Min Fairness Design for Energy-Efficient Quantized ISAC LEO Satellite Systems: \\ A Rate-Splitting Approach
\title{
A {Secure} Full-Duplex Wireless Circulator enabled by Non-Reciprocal Beyond-Diagonal RIS
% Max-Min Fair Energy-Efficient Beam Design\\ for Quantized ISAC LEO Satellite Systems:\\ A Rate-Splitting Approach
% {\footnotesize \textsuperscript{*}Note: Sub-titles are not captured in Xplore and
% should not be used}
% \thanks{Identify applicable funding agency here. If none, delete this.}
}

\author{
Ziang Liu
        and~Bruno Clerckx,~\IEEEmembership{Fellow,~IEEE}
\thanks{Z. Liu, and B. Clerckx are with the Communications \& Signal Processing (CSP) Group at the Dept. of Electrical and Electronic Engg., Imperial College London, SW7 2AZ, UK. (e-mails:\{ziang.liu20, b.clerckx\}@imperial.ac.uk).
}}
% \thanks{B. Clerckx is also with Silicon Austria Labs (SAL), Graz A-8010, Austria.}
% <-this % stops a space
% \thanks{J. Doe and J. Doe are with Anonymous University.}% <-this % stops a space
%\thanks{Manuscript received April 19, 2005; revised August 26, 2015.}

% \author{\IEEEauthorblockN{1\textsuperscript{st} Given Name Surname}
% \IEEEauthorblockA{\textit{dept. name of organization (of Aff.)} \\
% \textit{name of organization (of Aff.)}\\
% City, Country \\
% email address or ORCID}
% \and
% \IEEEauthorblockN{2\textsuperscript{nd} Given Name Surname}
% \IEEEauthorblockA{\textit{dept. name of organization (of Aff.)} \\
% \textit{name of organization (of Aff.)}\\
% City, Country \\
% email address or ORCID}
% \and
% \IEEEauthorblockN{3\textsuperscript{rd} Given Name Surname}
% \IEEEauthorblockA{\textit{dept. name of organization (of Aff.)} \\
% \textit{name of organization (of Aff.)}\\
% City, Country \\
% email address or ORCID}
% \and
% \IEEEauthorblockN{4\textsuperscript{th} Given Name Surname}
% \IEEEauthorblockA{\textit{dept. name of organization (of Aff.)} \\
% \textit{name of organization (of Aff.)}\\
% City, Country \\
% email address or ORCID}
% \and
% \IEEEauthorblockN{5\textsuperscript{th} Given Name Surname}
% \IEEEauthorblockA{\textit{dept. name of organization (of Aff.)} \\
% \textit{name of organization (of Aff.)}\\
% City, Country \\
% email address or ORCID}
% \and
% \IEEEauthorblockN{6\textsuperscript{th} Given Name Surname}
% \IEEEauthorblockA{\textit{dept. name of organization (of Aff.)} \\
% \textit{name of organization (of Aff.)}\\
% City, Country \\
% email address or ORCID}
% }

\maketitle

\begin{abstract}
Beyond-diagonal reconfigurable intelligent surface (BD-RIS) has arisen as a promising technology for enhancing wireless communication systems by enabling flexible and intelligent wave manipulation. This is achieved through the interconnections among the ports of the impedance network, enabling wave reconfiguration when they flow through the surface. Thus, the output wave at one port depends on waves impinging on neighboring ports, allowing non-local control of both phase and magnitude. Non-reciprocal (NR)-BD-RIS further enhances this capability by breaking circuit reciprocity and, consequently, channel reciprocity. 
% This feature potentially benefits communication among non-aligned transceivers. 
% \textcolor{red}{This feature of the NR-BD-RIS enables uni-directional communications where UE 1 transmits to UE 2 and UE 2 transmits to UE 3. In contrast, conventional reciprocal (R)-BD-RIS and diagonal (D)-RIS are constrained by circuit and channel reciprocity, thus they only allow bi-directional communications, where signal transmitted by one UE would flow back to previous UE.}
{In contrast to conventional reciprocal (R)-BD-RIS and diagonal (D)-RIS that are constrained by circuit and channel reciprocity such that they only allow bidirectional communications, \ie $\text{UE}_1 \rightleftarrows \text{UE}_2$, NR-BD-RIS can additionally enable uni-directional communications, that is, $\text{UE}_1 \rightarrow \text{UE}_2 \rightarrow \text{UE}_3$, hence effectively enabling a wireless circulator.}
Specifically, this paper introduces a novel application of NR-BD-RIS in full-duplex (FD) wireless circulators, where multiple FD devices communicate via an NR-BD-RIS. This system is particularly beneficial for secure transmission, as it enforces one-way communication among FD devices, suppresses signal from all other users (UE), and thus prevents eavesdropping. In addition, a physics-compliant system model is considered by incorporating structural scattering, also known as specular reflection. By accounting for this effect, the advantages of NR-BD-RIS are further validated. Specifically, we formulate an sum-rate maximization problem and propose an iterative optimization algorithm that employs block coordinate descent (BCD) and penalty dual decomposition (PDD) methods. {Numerical evaluations illustrate that NR-BD-RIS outperforms conventional R-BD-RIS and D-RIS in terms of sum-rate and secrecy rate.}
\end{abstract}

\begin{IEEEkeywords}
BD-RIS, {eavesdropping}, FD wireless circulator, non-reciprocity, structural scattering, {secure transmission}
\end{IEEEkeywords}

\section{Introduction}
Reconfigurable intelligent surface (RIS) has emerged as a promising technique for future wireless communications \cite{latva2019key}. It has gained substantial interest in both academic research and the wireless industry due to its capability to intelligently manipulate electromagnetic wave propagation, thereby improving energy and spectrum efficiency \cite{wu2019intelligent, huang2019reconfigurable, liu2021reconfigurable}. RIS consists of a planar array of low-cost, energy-efficient passive reflecting elements with controllable phases. This enables the adjustment of incident, reflected, refracted, and scattered signals, facilitating the optimization of the propagation environment to enhance received signal power, sum-rate \cite{guo2020weighted}, and energy efficiency \cite{huang2019reconfigurable} in wireless systems. 
% RIS has been shown to be effective in supporting various communication applications. For example, RIS-assisted multiple-input multiple-output (MIMO) systems can significantly reduce hardware costs and energy consumption \cite{wu2019intelligent}. Moreover, RIS is particularly advantageous for extending coverage in mmWave and THz communications \cite{xue2024survey}, which are prone to blockages. 
On the industrial front, RIS-related technologies, such as network-controlled repeaters (NCRs), have been introduced in the 3rd Generation Partnership Project (3GPP)'s Release-18 to enhance spectral efficiency \cite{3gpp2023repeater}. {Concurrently, with the increasing number of wireless devices and dynamic channel environments, physical layer (PHY) security has become a critical concern. Recent studies have investigated the potential of RIS to enable secure communication and against jamming and eavesdropping attacks\cite{9779086, niu2022joint}. }

A key limitation of conventional RISs \cite{guo2020weighted, huang2019reconfigurable, xue2024survey, wu2019intelligent, liu2021reconfigurable} is their inability to control the magnitude of impinging waves, as they can only adjust the phase. This restriction reduces the RIS's effectiveness in passive beamforming and electromagnetic wave manipulation. The reason for this is that, in theory, the RIS is equivalent to an array of scattering elements connected to a reconfigurable impedance network \cite{shen_modeling_2022, li_reconfigurable_2024, pozar_microwave_2021}. In conventional diagonal RIS (D-RIS), each port of the impedance network is grounded via its impedance component, resulting in a single-connected structure. 
% The limitation stems from the RIS architecture, which consists of multiple scattering elements connected to a reconfigurable impedance network \cite{shen_modeling_2022, li_reconfigurable_2024, pozar_microwave_2021}. In conventional RIS, each port of the impedance network is connected to its own impedance component to ground, resulting in a single-connected structure. 
The resulting scattering matrix is diagonal, leading to the term D-RIS. To address this limitation and enable control over both the phase and magnitude of impinging waves, \cite{shen_modeling_2022, li_reconfigurable_2024} developed beyond-diagonal RIS (BD-RIS). In BD-RIS, the ports of the impedance network are interconnected, allowing waves to propagate across the surface and providing greater flexibility for analog-domain wave manipulation. If all ports are interconnected, it is referred to as fully-connected BD-RIS \cite{shen_modeling_2022}, characterized by a full scattering matrix that offers maximum control over wave manipulation at the expense of increased hardware complexity. To trade off flexibility and hardware complexity, group-connected RIS architectures have been designed, where ports are grouped, and all ports within a group are interconnected. The scattering matrix in this case is block diagonal, offering more flexibility than D-RIS while being less complex than fully-connected RIS. To further optimize the balance between performance and hardware complexity, \cite{nerini2024beyond, nerini2023pareto} designed tree- and forest-connected structures and their performance-hardware complexity Pareto frontier is analyzed. These architectures enhance communication performance compared to conventional D-RIS. For example, in \cite{nerini2023closed, bjornson2024capacity}, closed-form solutions are used to maximize received power and capacity, demonstrating the superior performance of BD-RIS.

\begin{figure}[t]
    \centering
    \includegraphics[width= 0.4\linewidth]{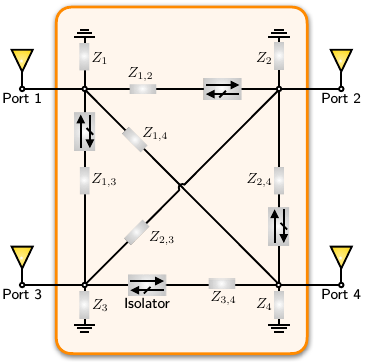}
    \caption{Illustration of a 4-port non-reciprocal BD-RIS (NR-BD-RIS) hardware implementation \cite{xu2024non}.}
    \label{fig:nr_hardware}
\end{figure}

Losslessness and reciprocity are two commonly assumed properties of the reconfigurable impedance network in BD-RIS, as in most existing works \cite{shen_modeling_2022, li_reconfigurable_2024, nerini2024beyond, nerini2023pareto, nerini2023closed, bjornson2024capacity}. Losslessness ensures that the network does not dissipate energy, meaning the power of the reflected waves equals that of the incident waves. This property is mathematically represented by the scattering matrix being unitary \cite{pozar_microwave_2021}. Reciprocity implies that the impedance between any two ports is symmetric, \ie the impedance from port $i$ to port $j$ is identical to that from port $j$ to port $i$. This is mathematically described by the scattering matrix being symmetric \cite{pozar_microwave_2021}. Reciprocal BD-RIS (R-BD-RIS) has been extensively studied in prior works \cite{shen_modeling_2022, nerini2024beyond, nerini2023pareto, nerini2023closed, bjornson2024capacity}. In fact, a new degree of freedom for manipulating the electromagnetic environment can be achieved by breaking the reciprocity of the impedance network. Thus, the symmetric constraint on the scattering matrix can be relaxed, leading to a non-reciprocal BD-RIS (NR-BD-RIS). Non-reciprocity can be realized by integrating non-reciprocal components such as radio isolators (cf. \fig{fig:nr_hardware}), circulators, and radio frequency couplers \cite{pozar_microwave_2021}. Additionally, \cite{li_reconfigurable_2022} proposed using switches based on RF micro-electromechanical systems (MEMS) \cite{rebeiz2001rf} to realize a non-diagonal phase shift matrix. Metasurfaces have also been explored for achieving non-reciprocity \cite{zhang2019breaking, zhang2018space, li2018metasurfaces, liaskos2020internet}. These studies indicate the feasibility of hardware implementation of NR-BD-RIS. Non-reciprocity has been leveraged in some works to enhance communication system performance. For example, \cite{li_reconfigurable_2022} demonstrated that a non-diagonal phase shift matrix can maximize channel gain. Similarly, \cite{rusek2024spatially} showed that a non-reciprocal scattering matrix can achieve spatial selectivity, mitigating interference from specific directions. Applications of NR-BD-RIS in secure communications and channel reciprocity attacks (CRACK) have been explored in \cite{pan_full_duplex_2021, wang2024channel}, respectively. Despite these advancements, the unique benefits of NR-BD-RIS as an enabler of new use cases not achievable with conventional D-RIS remain unexplored. This motivates further investigation into its potential for improving wireless communication systems.

\begin{figure}[t]
    \centering
    \includegraphics[width = 0.48\textwidth]{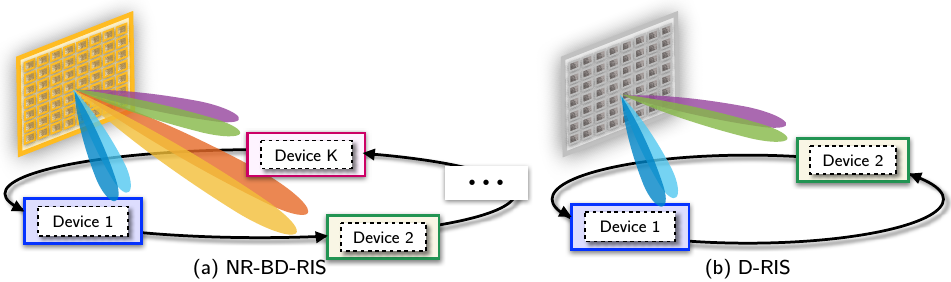}
    \centering
    \caption{A full-duplex (FD) wireless circulator enabled by (a) non-reciprocal BD-RIS (NR-BD-RIS). All FD devices communicate in one direction. {(b) The conventional D-RIS can only enable bi-directional communications where signal transmitted by device 2 would flow back to device 1. This is not a wireless circulator.}}
    \label{fig:circulator}
\end{figure}

Recent studies have demonstrated the benefits of NR-BD-RIS in full-duplex (FD) wireless communication systems. For instance, \cite{li_non-reciprocal_2024} theoretically shows that non-reciprocity improves received power for non-aligned downlink and uplink users (UE) in single-antenna FD systems, outperforming the performance of R-BD-RIS and D-RIS. Furthermore, \cite{liu2024non} extends this analysis to a more general FD case, where a base station (BS) with multiple antennas transmits data to multiple downlink and uplink UEs. The findings reveal that NR-BD-RIS achieves higher sum-rate performance, especially when downlink and uplink UEs are non-aligned. Building on these studies, we propose a novel NR-BD-RIS-enabled secure wireless circulator, which incorporates both direct and reflected links. {The secure wireless circulator is a secure system that can effectively suppress interference from unwanted FD devices and enforce one-way secure communication as illustrated in \fig{fig:circulator}(a).} {This wireless circulator can be achieved by NR-BD-RIS due to its ability to break the channel reciprocity and enable uni-directional communications where device 1 transmits to device 2 and device 2 transmits to device 3.
In contrast, the conventional D-RIS, as shown in \fig{fig:circulator}(b), is limited by hardware and channel reciprocity and only enable bi-directional communications, where signal transmitted by device 2 would flow back to device 1. This extra degrees of freedom enable NR-BD-RIS to manipulate waves and shape radio propagation with more flexibility, consequently leading to a better spectrum use, a higher sum-rate and a higher secrecy rate.}
% {The one-way manner of the proposed wireless circulator is particularly beneficial for physical layer security, as the UE can only receive information from the previous UE, preventing eavesdropping by other UEs.} 

\begin{figure}[t]
    \centering
    \includegraphics[width = 0.48\textwidth]{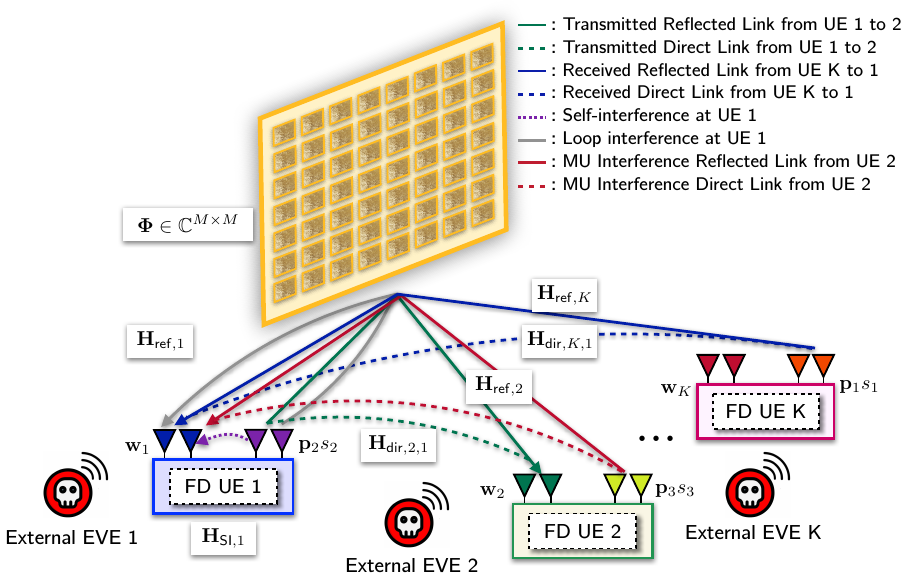}
    \centering
    \caption{{The system model of NR-BD-RIS enabled FD wireless circulator with external eavesdroppers (EVE).}}
    \label{fig:system}
\end{figure}

\bpara{Contributions and Overview of Results.} The contributions are summarized as follows: 
% In such setup, each FD user can transmit and receive simultaneously, and the wireless circulator allows signals to propagate in a single direction; the sent data from the $(k-1)^{\rm{th}}$ FD user is received by the $k^{\rm{th}}$ FD user. 
\begin{enumerate}[leftmargin = *,label =$\bullet$]
    \item We formulate a novel system model for the NR-BD-RIS-enabled secure FD wireless circulator in \fig{fig:system}, where multiple FD UEs communicate via NR-BD-RIS and signals are transmitted in one direction. {The signals leaked from other FD UEs (\ie multi-user (MU) interference or the eavesdropping signal) are suppressed, which prevents potential eavesdropping.} Each FD UE is equipped with multiple antennas for simultaneous transmission and reception. The model accounts for self-interference, loop interference, and multi-user interference. Both direct and reflected channels are included to illustrate the advantages of the NR-BD-RIS.
                        
    \item Unlike existing studies that neglect structural scattering, which represents specular reflection caused by the BD-RIS when inactive, we adopt a physics-compliant model that incorporates structural scattering. This model effectively demonstrates the consistent superiority of NR-BD-RIS over R-BD-RIS and D-RIS in terms of sum-rate performance.
                        
    \item We formulate an  sum-rate maximization problem for the NR-BD-RIS-enabled FD wireless circulator. The problem involves optimizing precoders, combiners of FD UEs, and the scattering matrix of the BD-RIS, using an iterative optimization approach. Specifically, a block coordinate descent (BCD) framework is adopted to iteratively optimize all variables, while the scattering matrix is optimized using the penalty dual decomposition (PDD) method. Compared to \cite{liu2024non}, this approach is designed for the FD wireless circulator scenario, incorporates the specific sum-rate maximization problem, and effectively handles symmetry, asymmetry, and diagonality constraints.
                    
    \item We demonstrate the unique benefit of NR-BD-RIS in effectively supporting more than two impinging and reflecting directions, which cannot be achieved by R-BD-RIS and conventional D-RIS. This unique feature enables the secure FD wireless circulator, facilitating one-way communication among multiple FD UEs. This benefit is validated through simulation results, which show that NR-BD-RIS consistently outperforms R-BD-RIS and D-RIS in terms of sum-rate and sum secrecy rate performances. This performance gain is attributed to the additional degree of freedom provided by non-reciprocity, which breaks channel reciprocity. The performance gain of NR-BD-RIS increases with the number of RIS elements, the group size of the BD-RIS, the number of antennas at each FD UE, and the number of FD UEs. Additionally, structural scattering enhances sum-rate performance by increasing channel gain. {The sum secrecy rate in the presence of internal and external EVEs, power of eavesdropping signals and the beampatterns are analyzed to demonstrate the secure transmission capability of the proposed system.}

\end{enumerate}

\bpara{Organization of This Paper.} 
% The organization of the paper is as follows. 
The system model is described in Section \ref{sec:sys}. Based on it, we formulate the weighted  sum-rate maximization problem in Section \ref{sec:pro}. Then, the optimization problem is transformed into a more tractable form. Subsequently, the proposed solution algorithm is presented in Section \ref{sec:algo}. In this section, the convergence and computational complexity are analyzed. The numerical evaluation is provided in Section \ref{sec:simu}. We summarize the paper in Section \ref{sec:con}.

\bpara{Notation.} The sets of binary, integer, real, and complex numbers are denoted as $\mathbb{B}$, $\mathbb{Z}$, $\mathbb{R}$, and $\mathbb{C}$, respectively. Bold uppercase letters represent matrices, bold lowercase letters denote vectors, and scalars are written in regular font. The real part of a complex number is expressed as $\Re(\cdot)$. For a matrix $\mathbf{X}$, its complex conjugate, transpose, Hermitian transpose, and inverse are represented by $\mathbf{X}^*$, $\mathbf{X}^\top$, $\mathbf{X}^H$, and $\mathbf{X}^{-1}$, respectively. The element located at the $i^{\rm{th}}$ row and $j^{\rm{th}}$ column of $\mathbf{X}$ is denoted as $\mathbf{X}(i,j)$. The identity matrix and the zero matrix are denoted by $\mathbf{I}$ and $\mathbf{0}$, respectively. The operations of vectorization, diagonal matrix, block diagonal matrix, trace operation, and Kronecker product are denoted by $\operatorname{vec}(\cdot)$, $\operatorname{diag}(\cdot)$, $\operatorname{blkdiag}(\cdot)$, $\Tr(\cdot)$, and $\otimes$, respectively.

\section{System model}
\label{sec:sys}
As illustrated in \fig{fig:system}, a secure FD wireless circulator is enabled by an $M$-element RIS. In this setup, all $K$ UEs operate in FD mode, allowing simultaneous transmission and reception. Each FD UE is equipped with $N_{t}$ transmit (TX) antennas and $N_{r}$ receive (RX) antennas, with $N_{t} = N_{r} = N$ for simplicity. {In addition, we consider both internal and external eavesdroppers (EVE). In the case of internal EVEs, the $k^{\rm{th}}$ FD UE can be eavesdropped by $i^{\rm{th}}$ FD UE, where $i \in \mathcal{K}, i \neq k, k+1$. For example, for the UE 1 in the circulator with three FD UEs, UE 1 itself cannot be eavesdropped and UE 2 receives wanted signal from UE 1, thus UE 1 can be eavesdropped by UE 3. In the case of external EVEs, we consider $K$ external EVE, each equipped with single antenna, attempting to intercept the communication among FD UEs.
The FD UEs and EVEs are both indexed by $\mathcal{K} = \{1, \dots, K \}$.} 

The following assumptions are made: 
\textit{i}) {Each FD UE has perfect instantaneous channel state information (CSI) with details in Remark \ref{remark1}.}
% The CSI is acquired using BD-RIS channel estimation techniques \cite{li_channel_2024}.
\textit{ii}) The $(k-1)^{\rm{th}}$ FD UE transmits data to the $k^{\rm{th}}$ FD UE. For $k=K$, the $K^{\rm{th}}$ FD UE transmits to the $1^{\rm{st}}$ FD UE, as depicted in \fig{fig:system}. 
{\textit{iii}) The UE-EVE pair is assumed in the scenario of external EVEs, \ie the $k^{\rm{th}}$ FD UE is eavesdropped by the $k^{\rm{th}}$ EVE \cite{bloch2011physical, niu2021weighted}. }
Therefore, the FD UEs are connected in a circular manner, allowing signals to propagate in one direction. This configuration is referred to as a wireless circulator.
We use $\mathbf{\Phi} \in \mathbb{C}^{M \times M}$ to represent the scattering matrix of the BD-RIS. 
In the case of a group-connected BD-RIS, the group size is denoted as $M_g$, and the total number of groups is $G = M/M_g$. The group-connected BD-RIS is expressed as $\mathbf{\Phi} = \operatorname{blkdiag}(\mathbf{\Phi}_1, \cdots, \mathbf{\Phi}_G)$.
% \begin{equation}
%     \mathbf{\Phi} = \operatorname{blkdiag}(\mathbf{\Phi}_1, \cdots, \mathbf{\Phi}_G).
% \end{equation}
Notably, the D-RIS and fully-connected BD-RIS are special cases where $G = M$ and $G = 1$, respectively.

\begin{figure}[t]
    \centering
    \includegraphics[width = 0.35\textwidth]{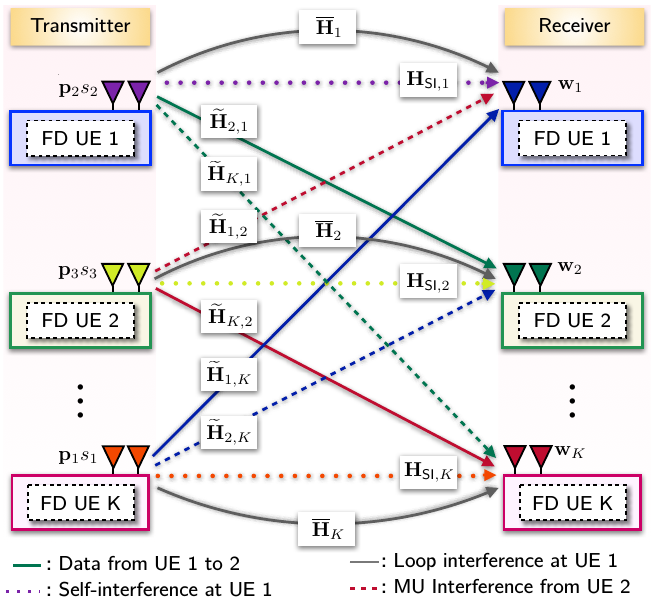}
    \centering
    \caption{Effective channels in the NR-BD-RIS enabled FD wireless circulator.}
    \label{fig:channels}
\end{figure}

Consider the $(k-1)^{\rm{th}}$ FD UE as an example. During data transmission, the FD UE applies the precoding matrix $\mathbf{p}_k \in \mathbb{C}^{N}$ to the transmit symbol $s_k$, which is intended for the $k^{\rm{th}}$ FD UE. Subsequently, the symbol is processed in baseband and up-converted. The resulting signal is sent via $N$ antennas and propagates through both the direct and reflected paths. At $k^{\rm{th}}$ FD UE, received signals are processed using the combiner $\mathbf{w}_k \in \mathbb{C}^{N}$. Let $\mathbf{P} \triangleq [\mathbf{p}_1, \cdots, \mathbf{p}_k] \in \mathbb{C}^{N \times K}$ and $\mathbf{W} \triangleq [\mathbf{w}_1, \cdots, \mathbf{w}_k] \in \mathbb{C}^{N \times K}$ represent the precoders and combiners for all $K$ FD UEs. Note that the transpose operator $(\cdot)^\top$ is used in channel-related expressions instead of the Hermitian operator $(\cdot)^H$ to more accurately model channel reciprocity. 
As illustrated in \fig{fig:channels}, the effective channels are defined as $\widetilde{\mathbf{H}}_{k, i} \triangleq \mathbf{H}^\top_{\mathsf{dir},k, i} + \mathbf{H}_{\mathsf{ref},k}^\top (\mathbf{\Phi}-\mathbf{I} ) \mathbf{H}_{\mathsf{ref},i}  \in \mathbb{C}^{N \times N}$ and $\overline{\mathbf{H}}_k = \mathbf{H}_{\mathsf{ref},k}^\top (\mathbf{\Phi-\mathbf{I}} ) \mathbf{H}_{\mathsf{ref},k}\in \mathbb{C}^{N \times N}$. Here, $\mathbf{H}_{\mathsf{dir},k, i} \in \mathbb{C}^{N \times N}$ represents the direct channel between the $k^{\rm{th}}$ and $i^{\rm{th}}$ FD UEs, while $\mathbf{H}_{\mathsf{ref},k} \in \mathbb{C}^{M \times N}$ denotes the channel between the $k^{\rm{th}}$ FD UE and the RIS. 

% {
% The $j^{\rm{th}}$ EVE receives signal transmitted from the $k^{\rm{th}}$ FD UE. The effective channel from the $k^{\rm{th}}$ FD UE to the $j^{\rm{th}}$ EVE is defined as ${\mathbf{g}}_{\mathsf{e},j,k} \triangleq \mathbf{h}^\top_{\mathsf{dir, e},j,k} + \mathbf{h}_{\mathsf{ref, e},j}^\top (\mathbf{\Phi}-\mathbf{I} ) \mathbf{H}_{\mathsf{ref},k}  \in \mathbb{C}^{1 \times N}$, where $\mathbf{h}_{\mathsf{dir,e},j,k} \in \mathbb{C}^{N \times 1}$ represents the direct channel between the $k^{\rm{th}}$ FD UE and the $j^{\rm{th}}$ EVE, while $\mathbf{h}_{\mathsf{ref,e},j} \in \mathbb{C}^{M \times 1}$ denotes the channel between the RIS and the $j^{\rm{th}}$ EVE.
% }
{
The $k^{\rm{th}}$ external EVE receives signal transmitted from the $k^{\rm{th}}$ FD UE but also other FD UEs. Thus, the effective channel from the $j^{\rm{th}}$ FD UE, where $j\in \mathcal{K}$, to the $k^{\rm{th}}$ EVE is defined as $\widetilde{\mathbf{g}}_{k,j} \triangleq \mathbf{g}^\top_{\mathsf{dir},k,j} + \mathbf{g}_{\mathsf{ref},k}^\top (\mathbf{\Phi}-\mathbf{I} ) \mathbf{H}_{\mathsf{ref},j}  \in \mathbb{C}^{1 \times N}$, where $\mathbf{g}_{\mathsf{dir},k,j} \in \mathbb{C}^{N \times 1}$ represents the direct channel between the $j^{\rm{th}}$ FD UE and the $k^{\rm{th}}$ EVE, while $\mathbf{g}_{\mathsf{ref},j} \in \mathbb{C}^{M \times 1}$ denotes the channel between the RIS and the $k^{\rm{th}}$ EVE.
}

\subsection{Structural Scattering} 
The terms $- \mathbf{H}_{\mathsf{ref},k}^\top \mathbf{H}_{\mathsf{ref},i}$ and $- \mathbf{H}_{\mathsf{ref},k}^\top \mathbf{H}_{\mathsf{ref},k}$ represent structural scattering, which is also termed as specular reflection caused by the RIS when it is inactive. It is worth noting that the commonly used model that ignores structural scattering can be derived from this model by omitting the $-\mathbf{I}$ term immediately following $\mathbf{\Phi}$ in all channel expressions. Although structural scattering is often neglected in most RIS studies, it is more physics-compliant and can significantly impact system performance \cite{hansen1989relationships, abrardo2024design, li_non-reciprocal_2024, nerini2024physics}. Specifically, the relationship between the impedance and scattering matrix is given by \cite{nerini2024universal}
\begin{equation}
\mathbf{S}_{R I}=\frac{\mathbf{Z}_{R I}}{2 Z_0}, \quad \mathbf{S}_{I T}=\frac{\mathbf{Z}_{I T}}{2 Z_0},
\end{equation}
\vspace{-5pt}
\begin{equation}
\mathbf{S}_{R T}=\frac{1}{2 Z_0}\left(\mathbf{Z}_{R T}-\frac{\mathbf{Z}_{R I} \mathbf{Z}_{I T}}{2 Z_0}\right).
\label{eq:89}
\end{equation}
The transmission impedance matrices from the transmitter to receiver, from the transmitter to RIS, and from the RIS to receiver are denoted by $\mathbf{Z}_{R T} \in \mathbb{C}^{N \times N}$, $\mathbf{Z}_{I T} \in \mathbb{C}^{M \times N}$,  and $\mathbf{Z}_{R I} \in \mathbb{C}^{N \times M}$, respectively. $Z_0$ is the characteristic impedance used to compute the S-parameters, typically equal to $Z_0 = 50\, \Omega$. Similarly, $\mathbf{S}_{R T} \in \mathbb{C}^{N \times N}$, $\mathbf{S}_{I T} \in \mathbb{C}^{M \times N}$,  and $\mathbf{S}_{R I} \in \mathbb{C}^{N \times M}$, denote the transmission scattering matrices from the transmitter to receiver, from the transmitter to RIS, and from the RIS to receiver, respectively.

In most existing RIS literature, scattering matrices between devices are regarded as wireless channels such that the following relationship is commonly used:
% \vspace{-2pt}
\begin{equation}
\mathbf{H}_{R T}=\mathbf{S}_{R T}, \quad \mathbf{H}_{R I}=\mathbf{S}_{R I}, \quad \mathbf{H}_{I T}=\mathbf{S}_{I T}.
\label{eq:90}
\end{equation}
% \vspace{-2pt}
In addition, the approximation of \eqref{eq:89} that $\mathbf{S}_{R T} \approx \frac{\mathbf{Z}_{R T}}{2 Z_0}$ is assumed, thus the channel model can be expressed as 
% \vspace{-2pt}
\begin{equation}
\mathbf{H}=\mathbf{H}_{R T}+\mathbf{H}_{R I} \boldsymbol{\Theta} \mathbf{H}_{I T}.
\label{eq:wsschannel}
\end{equation}
% \vspace{-2pt}

However, the direct mapping from the scattering parameter to wireless channels is not always physically compliant. In contrast, since the impedance parameters $\mathbf{Z}_{R T}, \mathbf{Z}_{R I}$, and $\mathbf{Z}_{IT}$ capture the open-circuit radiation patterns between devices, they accurately characterize the wireless channels. With \eqref{eq:89}, the channel model is more accurately expressed as 
% \vspace{-2pt}
\begin{equation}
    \mathbf{H}=\mathbf{H}_{R T}+\mathbf{H}_{R I} (\boldsymbol{\Theta} - \mathbf{I}) \mathbf{H}_{I T}.
    \label{eq:sschannel}
\end{equation}
% \vspace{-2pt}

Comparing  \eqref{eq:wsschannel} and \eqref{eq:sschannel}, we observe an additional term $-\mathbf{H}_{R I} \mathbf{I} \mathbf{H}_{I T}$, which refers to the structural scattering at RIS. In this sense, if we consider a completely obstructed direct channel, \ie $\mathbf{Z}_{R T} = \mathbf{0}$, and turn the RIS OFF, \ie $\boldsymbol{\Theta} = \mathbf{0}$, then we have $\mathbf{H}=-\mathbf{H}_{R I} \mathbf{H}_{I T}$ instead of being zero. 
This term exists since $\boldsymbol{\Theta} = \mathbf{0}$ is achieved by $\mathbf{Z}_I = Z_0\mathbf{I}$ with $\mathbf{Z}_I \in \mathbb{C}^{M \times M}$ denoting the impedance matrix of the $M$-port reconfigurable impedance network. This means that  RIS still radiates due to the non-zero current in RIS elements. We include structural scattering in our model to better understand its impact and highlight advantages of NR-BD-RIS.

% In our paper, we use $- \mathbf{H}_{\mathsf{ref},k}^\top \mathbf{H}_{\mathsf{ref},i}$ and $- \mathbf{H}_{\mathsf{ref},k}^\top \mathbf{H}_{\mathsf{ref},k}$ to model structural scattering to better understand its impact and highlight the performance advantages of NR-BD-RIS.

\subsection{Signal Model}
With such accurate model, the received signal at the $k^{\rm{th}}$ FD UE is given by
% \begin{equation}
%     \begin{aligned}
%     & y_{k} = \textcolor{black}{\mathbf{w}_k^H \widetilde{\mathbf{H}}_{k, k-1} \mathbf{p}_k s_k } + \underbrace{ \mathbf{w}_k^H \mathbf{H}_{\mathsf{SI},k} \mathbf{p}_{k+1} s_{k+1}}_{\text{Self-interference}} \\
%     & + \textcolor{black}{\underbrace{ \mathbf{w}_k^H \overline{\mathbf{H}}_k \mathbf{p}_{k+1} s_{k+1}}_{\text{Loop Interference}}} \!\!\!+\!\!\! \underbrace{ \sum_{i \in \mathcal{K}, i \neq k, k+1 } \!\!\!\!\!\! \mathbf{w}_k^H \widetilde{\mathbf{H}}_{k, i-1} \mathbf{p}_i s_i}_{\text{MU Interference}}   +\mathbf{w}_k^H \mathbf{n}_k,
%     \label{eq:dlsignal}
%     \end{aligned}
% \end{equation}
\begin{equation}
    \begin{aligned}
     y_{k} & = \textcolor{black}{\mathbf{w}_k^H \widetilde{\mathbf{H}}_{k, k-1} \mathbf{p}_k s_k } + \underbrace{ \mathbf{w}_k^H \mathbf{H}_{\mathsf{SI},k} \mathbf{p}_{k+1} s_{k+1}}_{\text{Self-interference}} \\
    & + \textcolor{black}{\underbrace{ \mathbf{w}_k^H \overline{\mathbf{H}}_k \mathbf{p}_{k+1} s_{k+1}}_{\text{Loop Interference}}} + \textcolor{black}{\underbrace{ \sum_{i \in \mathcal{K}, i \neq k, k+1 }  \mathbf{w}_k^H \widetilde{\mathbf{H}}_{k, i-1} \mathbf{p}_i s_i}_{\text{MU Interference}} }\\
    & +\mathbf{w}_k^H \mathbf{n}_k,  \,  \forall k \in \mathcal{K},
    \label{eq:dlsignal}
    \end{aligned}
\end{equation}
% \begin{equation}
% \begin{aligned}
% & y_{k} = \mathbf{w}_k^H \mathbf{H}_{k}^\top (\mathbf{\Phi} \! - \! \mathbf{I}) \mathbf{H}_{k-1} \mathbf{p}_k s_k 
% \! \\
% & + \underbrace{ \mathbf{w}_k^H \mathbf{H}_{\mathsf{SI},k} \mathbf{p}_{k+1} s_{k+1}}_{\text{Self-interference}} +   \underbrace{ \mathbf{w}_k^H \mathbf{H}_{k}^\top \! (\mathbf{\Phi} \! - \! \mathbf{I}) \mathbf{H}_{k} \mathbf{p}_{k+1} s_{k+1}}_{\text{Loop Interference}}\\
% & + \underbrace{ \sum_{i \in \mathcal{K}, i \neq k, k+1 }  \mathbf{w}_k^H \mathbf{H}_{k}^\top (\mathbf{\Phi} - \mathbf{I}) \mathbf{H}_{i-1} \mathbf{p}_i s_i}_{\text{MU Interference}} 
% % & + \underbrace{\sqrt{P_u} \mathbf{w}_k^H \mathbf{H}_\mathsf{SI} \mathbf{p}_{k+1} s_{k+1}}_{\text{Self-interference}}\\
% +\mathbf{w}_k^H \mathbf{n}_k,  \,  \forall k \in \mathcal{K},
% \label{eq:dlsignal}
% \end{aligned}
% \end{equation}
where $\mathbf{H}_{\mathsf{SI},k} \in \mathbb{C}^{N \times N}$ represents the self-interference (SI) channel of the $k^{\rm{th}}$ FD UE. Specifically, $\mathbf{H}_{\mathsf{SI},k} = \mathbf{H}_{\mathsf{dir},k,k}$. The additive Gaussian white noise (AWGN) is represented by $\mathbf{n}_k \sim \mathcal{CN}(\mathbf{0}, \sigma^2 \mathbf{I}_N)$. {To prevent eavesdropping, we aim to suppress the signals from other UEs except $(k-1)^\mathrm{th}$ UE, which are referred to as MU interference. }

The signal-to-interference-plus-noise ratio (SINR) at the $k^{\rm{th}}$ FD UE is calculated as 
% \begin{equation}
%     \gamma_{k} =\frac{ |\mathbf{w}_k^H \widetilde{\mathbf{H}}_{k, k-1} \mathbf{p}_k |^2}{ I_k (\mathbf{w}_k, \{ \mathbf{p}_i \}_{i \neq k, k+1}, \mathbf{\Phi}) + \| \mathbf{w}_k \|^2_F \sigma^2},
% \label{eq:sinr}
% \end{equation}
\begin{equation}
    \gamma_{k} =\frac{ |\mathbf{w}_k^H \widetilde{\mathbf{H}}_{k, k-1} \mathbf{p}_k |^2}{ I_k + \| \mathbf{w}_k \|^2_F \sigma^2},
\label{eq:sinr}
\end{equation}
where the interference power term incorporating SI, loop interference, and MU interference is expressed by
\color{black}
\begin{equation}
\begin{aligned}
    I_k \triangleq \!\!\!\! \sum_{\substack{i \in \mathcal{K},\\ i \neq k, k+1}}  \!\!\!\!| \mathbf{w}_k^H \widetilde{\mathbf{H}}_{k,i-1} \mathbf{p}_{i}|^2 \!  +\! | \mathbf{w}_k^H \mathbf{H}_{\mathsf{SI},k} \mathbf{p}_{k+1}\! + \!\mathbf{w}_k^H \overline{\mathbf{H}}_k \mathbf{p}_{k+1}|^2.
\end{aligned}
\end{equation}
The achievable rate at the $k^{\rm{th}}$ FD UE is $R_{k} = \log_2 (1+\gamma_{k})$.
% \begin{equation}
%     R_{k} = \log_2 (1+\gamma_{k}),
% \label{eq:rate}
% \end{equation}

\subsection{Achievable Secrecy Rate in the Presence of Internal Eavesdroppers}
If EVEs are in the FD wireless circulator, the $k^{\rm{th}}$ FD UE can be eavesdropped by $i^{\rm{th}}$ FD UE, where $i \in \mathcal{K}, i \neq k, k+1$. 
% For example, for the UE 1 in the circulator with three FD UEs, UE 1 itself cannot be eavesdropped and UE 2 receives wanted signal from UE 1, thus UE 1 can be eavesdropped by UE 3. 
The received signal at EVE, \ie $i^{\rm{th}}$ FD UE, is given by
\begin{equation}
    \begin{aligned}
     y_{i,k}^{\mathsf{int}} & = \mathbf{w}_i^H \widetilde{\mathbf{H}}_{i, k} \mathbf{p}_{k+1} s_{k+1} 
     + {\underbrace{ \sum_{j \in \mathcal{K}, j \neq i-1, i, k}  \mathbf{w}_i^H \widetilde{\mathbf{H}}_{i, j} \mathbf{p}_{j+1} s_{j+1}}_{\text{Unwanted Eavesdropping Signal}} }  \\
    & + \underbrace{ \mathbf{w}_i^H \mathbf{H}_{\mathsf{SI},i} \mathbf{p}_{i+1} s_{i+1}}_{\text{Self-interference}} + {\underbrace{ \mathbf{w}_i^H \overline{\mathbf{H}}_i \mathbf{p}_{i+1} s_{i+1}}_{\text{Loop Interference}}}  + \underbrace{\mathbf{w}_i^H \widetilde{\mathbf{H}}_{i, i-1} \mathbf{p}_{i} s_{i}}_{\text{Data Signal}} \\
    & +\mathbf{w}_i^H \mathbf{n}_i,  \,  \forall k \in \mathcal{K},
    \label{eq:evesignal_internal}
    \end{aligned}
\end{equation}
Thus, the eavesdropping signal at the $i^{\rm{th}}$ FD UE for eavesdropping the $k^{\rm{th}}$ FD UE is $\mathbf{w}_i^H \widetilde{\mathbf{H}}_{i,k} \mathbf{p}_{k+1}$ and the SINR at the $i^{\mathrm{th}}$ FD UE is expressed as
% \begin{equation}
%     \gamma_{i,k}^{\mathsf{int}} =\frac{ |\mathbf{w}_i^H \widetilde{\mathbf{H}}_{i, k} \mathbf{p}_{k+1} |^2}{ I_i^\mathsf{int} (\mathbf{w}_i, \{ \mathbf{p}_k \}_{k \neq i-1, i}, \mathbf{\Phi}) + \| \mathbf{w}_i \|^2_F \sigma^2},
% \label{eq:sinr_internal_eve}
% \end{equation}
\begin{equation}
    \gamma_{i,k}^{\mathsf{int}} = \frac{ |\mathbf{w}_i^H \widetilde{\mathbf{H}}_{i, k} \mathbf{p}_{k+1} |^2}{ I_i^\mathsf{int}  + \| \mathbf{w}_i \|^2_F \sigma^2},
\label{eq:sinr_internal_eve}
\end{equation}
where the interference power term at the $i^{\rm{th}}$ FD UE is 
% \begin{equation}
% \begin{aligned}
%     I_i^\mathsf{int} & =  \sum_{j \in \mathcal{K}, j \neq k-1, k, k+1}  | \mathbf{w}_i^H \widetilde{\mathbf{H}}_{i,j} \mathbf{p}_{j+1}|^2 \\
%     & + | \mathbf{w}_i^H \mathbf{H}_{\mathsf{SI},i} \mathbf{p}_{i+1} + \mathbf{w}_i^H \overline{\mathbf{H}}_i \mathbf{p}_{i+1}|^2.
% \end{aligned}
% \end{equation}
\begin{equation}
\begin{aligned}
    I_i^\mathsf{int} = \!\!\!\! \sum_{\substack{j \in \mathcal{K},\\ j \neq i, k}} \!\!\!\! | \mathbf{w}_i^H \widetilde{\mathbf{H}}_{i,j} \mathbf{p}_{j+1}|^2 \! + \! | \mathbf{w}_i^H \mathbf{H}_{\mathsf{SI},i} \mathbf{p}_{i+1} \! + \! \mathbf{w}_i^H \overline{\mathbf{H}}_i \mathbf{p}_{i+1}|^2.
\end{aligned}
\end{equation}
The achievable rate at the $i^{\rm{th}}$ FD UE for eavesdropping the $k^{\rm{th}}$ FD UE is $R_{i,k}^\mathsf{int} = \log_2 (1+\gamma_{i,k}^\mathsf{int})$. Thus, the sum rate of all internal EVEs for the $k^{\rm{th}}$ FD UE is $R_{\mathsf{sum},k}^\mathsf{int} = \sum_{i \in \mathcal{K}, i \neq k, k+1} R_{i,k}^\mathsf{int}$.

Consequently, the secrecy rate in the presence of internal EVEs of the $k^{\rm{th}}$ FD UE is defined by \cite{bloch2011physical}
\begin{equation}
    R_{\mathsf{s},k}^\mathsf{int} = \left[ R_{k} - R_{\mathsf{sum},k}^\mathsf{int} \right]^+,
\label{eq:secrecyrate_interal}
\end{equation}
where $[x]^+ = \max(x,0)$ guarantee non-negative secrecy rate.

% Thus, the sum rate for eavesdropping the $k^{\rm{th}}$ FD UE by internal eavesdroppers is 
% \begin{equation}
%     \sum_{i \in \mathcal{K}, i \neq k, k+1} R_{i}^\mathsf{int} = \log_2 (1+\gamma_{i,k}).
% \label{eq:internal_everate}
% \end{equation}

\subsection{Achievable Secrecy Rate in the Presence of External Eavesdroppers}
Another case is that there are external EVEs, which attempt to eavesdrop the FD wireless circulator. At the $k^{\rm{th}}$ EVE, the received signal is given by \cite{bloch2011physical, niu2021weighted}
\begin{equation}
    \begin{aligned}
     y_{k,k}^{\mathsf{ext}} & =  \widetilde{\mathbf{g}}_{k,k} \mathbf{p}_k s_k  + \underbrace{ \sum_{i \in \mathcal{K}, i \neq k}  \widetilde{\mathbf{g}}_{k,i} \mathbf{p}_i s_i}_{\text{Interference}} +{\mathbf{n}}_{\mathsf{e},k},  \,  \forall k \in \mathcal{K},
    \label{eq:everxsignal}
    \end{aligned}
\end{equation}
where $\mathbf{n}_{\mathsf{e},k} \sim \mathcal{CN}(0, \sigma^2_{\mathsf{e}})$ denotes the AWGN at the $k^{\rm{th}}$ EVE with noise variance $\sigma^2_{\mathsf{e}}$. 
The SINR at the $k^{\rm{th}}$ EVE is given by
\begin{equation}
    \gamma_{k}^\mathsf{ext} =\frac{ |\widetilde{\mathbf{g}}_{k,k} \mathbf{p}_k |^2}{ \sum_{i \in \mathcal{K}, i \neq k}  |\widetilde{\mathbf{g}}_{k,i} \mathbf{p}_i|^2 + \sigma^2_{\mathsf{e}}}.
\label{eq:sinre}
\end{equation}
% \subsection{Achievable Rate and Achievable Secrecy Rate}
% Based on the above SINR expressions, the achievable rate at the $k^{\rm{th}}$ FD UE is given by
% \begin{equation}
%     R_{k} = \log_2 (1+\gamma_{k}),
% \label{eq:rate}
% \end{equation}

% \subsubsection{Achievable Secrecy Rate of Internal Eavesdroppers}
% If we consider the eavesdroppers are in the FD wireless circulator, the $k^{\rm{th}}$ FD UE can be eavesdropped by $i \in \mathcal{K}, i \neq k-1, k$ FD UEs. For example, for the UE 1 in the circulator with three FD UEs, UE 1 itself cannot be eavesdropped and UE 1 receives wanted signal from UE 3, thus UE 1 can be eavesdropped by UE 2. Therefore the SINR 

% \subsubsection{Achievable Secrecy Rate of External Eavesdroppers}
The achievable rate at the $k^{\rm{th}}$ EVE is expressed as $R_{k}^\mathsf{ext} = \log_2 (1+\gamma_{k}^\mathsf{ext})$.
% \begin{equation}
%     R_{\mathsf{e},k} = \log_2 (1+\gamma_{\mathsf{e},k}).
% \label{eq:everate}
% \end{equation}
The secrecy rate in the presence of the $k^{\rm{th}}$ FD UE is defined as \cite{bloch2011physical, niu2021weighted}
\vspace{-3pt}
\begin{equation}
    R_{\mathsf{s},k}^\mathsf{ext} = \left[ R_{k} - R_{k}^\mathsf{ext} \right]^+.
\label{eq:secrecyrate}
\end{equation}
% where $[x]^+ = \max(x,0)$ guarantee non-negative secrecy rate. 

\begin{remark}
\label{remark1}
Perfect instantaneous CSI at the FD UEs is assumed to evaluate the performance upper bound for the proposed FD wireless circulator. Existing BD-RIS channel estimation methods \cite{li_channel_2024, wang2025low, samy2025low} can support practical deployment. In particular, \cite{li_channel_2024} provides an LS-based scheme with joint pilot and BD-RIS training design. In a circulator setup, the NR-BD-RIS enters a training mode with predefined scattering matrices, and the FD UEs transmit orthogonal pilots in separate slots to recover the cascaded channels via LS estimation. Subsequently, The NR-BD-RIS switches to the circulator mode and enables secure one-way communication.
\end{remark}
\color{black}
% 
% \begin{remark}
% \label{remark1_1}
% The ideal RIS model used in this work assumes continuous and lossless control of the scattering matrix. In practice, many factors constrain NR-BD-RIS hardware implementation \cite{li2025tutorial}. 
% 1) The impedance and admittance loads have discrete tuning steps rather than continuous values, and phase noise is also an issue. 
% 2) The interconnections and admittance components are lossy due to the interconnection loss between ports of the reconfigurable impedance network and the loss from the circuits of the reconfigurable admittance components. 
% 3) The impedance network exhibits frequency-dependent behaviour because the responses of the inductance and capacitance vary with the frequency of the applied signal. 
% 4) Mutual coupling among interconnected ports changes the effective response of each element. 
% 5) In addition, BD-RIS has higher circuit complexity since more impedance elements are needed in the interconnections among ports. New architectures such as tree- and forest-connected \cite{nerini2024beyond,nerini2023pareto} are proposed to achieve better trade-offs between performance and complexity. 
% These hardware constraints may introduce deviations from the ideal model, and incorporating such non-idealities into the system design is an important direction for future work.
% \end{remark}
% \color{black}

\vspace{-10pt}
\section{Problem Formulation and Transformation}
\label{sec:pro}
In this section, we formulate the weighted  sum-rate maximization problem for the secure FD wireless circulator. The optimization problem is then transformed using the fractional programming method into a more tractable form. In the optimization, we design the precoders and combiners of FD UEs, and the scattering matrix of the BD-RIS. 
\vspace{-10pt}
\subsection{Problem Formulation}
We aim to maximize the weighted sum-rate of all FD UEs:
\begin{equation}
    f_o(\mathbf{P}, \mathbf{W}, \mathbf{\Phi}) \triangleq  \sum_{k \in \mathcal{K}} \alpha_k \log_2 (1+\gamma_{k}).
    \label{eq:sumrate}
\end{equation}
In \eqref{eq:sumrate}, $\alpha_k$ denotes the weight for each FD UE, where the condition $\sum_{k \in \mathcal{K}} \alpha_k = 1$ should be satisfied. Subsequently, the optimization problem is given by 
\begin{maxi!}|s|[2]                   % mini! = minimize 
    {\mathbf{P}, \mathbf{W}, \mathbf{\Phi}}                               % optimization variable
    {f_o(\mathbf{P}, \mathbf{W}, \mathbf{\Phi}) \label{eq:op1}}   % objective function and label
    {\label{eq:p1}}             % label for opt problem
    {\mathcal{P}1:} 
    % optimization result
    {} 
    \addConstraint{\|\mathbf{p}_k\|^2_F}{\leq P_{k-1}, \quad \forall k \in \mathcal{K} \label{eq:op1c1}}
    \addConstraint{\|\mathbf{w}_k\|^2_F}{= 1, \quad \forall k \in \mathcal{K}, \label{eq:op1c2}}
    \addConstraint{\mathbf{\Phi}}{\in\mathcal{R}_i, \,  i \in\{1,2\}, \label{eq:op1c4}}
    \addConstraint{\mathbf{\Phi}}{\in \mathcal{S}_\ell, \, \ell \in \{ 1,2 \},  \label{eq:op1c3}}.
\end{maxi!}
% \begin{maxi!}|s|[2]                   % mini! = minimize 
%     {\mathbf{P}, \mathbf{W}, \mathbf{\Phi}}                               % optimization variable
%     {f_o(\mathbf{P}, \mathbf{W}, \mathbf{\Phi}) \label{eq:op1}}   % objective function and label
%     {\label{eq:p1}}             % label for opt problem
%     {\mathcal{P}1:} 
%     % optimization result
%     {} 
%     \addConstraint{\|\mathbf{p}_k\|^2_F}{\leq P_{k-1}, \quad \forall k \in \mathcal{K} \label{eq:op1c1}}
%     \addConstraint{\|\mathbf{w}_k\|^2_F}{= 1, \quad \forall k \in \mathcal{K}\label{eq:op1c2}}
%     \addConstraint{\mathbf{\Phi}}{\in\mathcal{R}_i, \quad \forall i \in\{1,2\}, \label{eq:op1c4}}
%     \addConstraint{\mathbf{\Phi}}{\in \mathcal{S}_\ell, \quad \forall \ell \in \{ 1,2,3 \} \label{eq:op1c3}}.
% \end{maxi!}
The transmit and receive power constraints are represented by \eqref{eq:op1c1} and \eqref{eq:op1c2}, respectively. The transmit power of the $(k-1)^{\rm{th}}$ FD UE for sending signals to the $k^{\rm{th}}$ FD UE is denoted by $P_{k-1}$.
{Constraint \eqref{eq:op1c4} specifies the symmetry of the scattering matrix: for R-BD-RIS, $\mathbf{\Phi}\in\mathcal{R}_1 = \{\mathbf{\Phi}  |\mathbf{\Phi} = \mathbf{\Phi}^\mathsf{T}\}$, and for NR-BD-RIS, $\mathbf{\Phi}\in\mathcal{R}_2 = \{\mathbf{\Phi}  | \mathbf{\Phi} \ne \mathbf{\Phi}^\mathsf{T}\}$. }
Constraint \eqref{eq:op1c3} enforces losslessness for both group-connected and fully-connected in the multi-port impedance network \cite{shen_modeling_2022, pozar_microwave_2021}, constraining $\mathbf{\Phi}$ as follows: \textit{i}) $\mathcal{S}_1 = \{\mathbf{\Phi} = \operatorname{blkdiag}(\mathbf{\Phi}_1, \cdots, \mathbf{\Phi}_G) | \boldsymbol{\Phi}_{g}^H \boldsymbol{\Phi}_{ g}= \mathbf{I}, \quad \forall g \in \mathcal{G} \}$ for group-connected BD-RIS, and \textit{ii}) $\mathcal{S}_2 = \{\mathbf{\Phi} |  \mathbf{\Phi}^H \mathbf{\Phi} = \mathbf{I} \}$ for fully-connected BD-RIS. We adopt a fractional programming-based approach \cite{shen2018fractional1} to reformulate the problem into a more tractable form. {Note that D-RIS has single-connected structure. Its constraint is $\mathbf{\Phi}\in\{ \mathbf{\Phi} = \text{diag}(\phi_1, \ldots,\phi_M) | |\phi_m | = 1, \, \forall m \in \mathcal{M} \}$. }
\vspace{-5pt}
% The problem $\mathcal{P}1$ is challenging due to the intractable $\log(\cdot)$ term, the fractional form in the objective function, and the non-convex constraints. To address this, inspired by \cite{shen2018fractional1, shen2018fractional2}, we adopt a fractional programming-based approach to reformulate the problem into a more tractable form, which is then solved iteratively.

% Problem $\mathcal{P}1$ includes an intractable $\log(\cdot)$ term and a fractional structure in the objective function, along with non-convex constraints. These conditions make it difficult to solve directly. To address this, we use a fractional programming-based method \cite{shen2018fractional1, shen2018fractional2} to make $\mathcal{P}1$ more tractable and subsequently solve it using an iterative method.

\subsection{Problem Transformation}
\vspace{-5pt}
\bpara{Lagrange Dual Transformation.}
To address the complexity of the fractional term, we decouple it from the $\log(\cdot)$ in the objective function \eqref{eq:op1} using the Lagrangian dual transformation \cite{shen2018fractional1}. Thus, we have a more manageable summation involving a new fractional term, expressed as
\begin{equation}
\begin{aligned}
    f_\iota(& \mathbf{P}, \mathbf{W}, \mathbf{\Phi})  = \\
    & \sum_{k \in \mathcal{K}} \alpha_k \Bigg( \!  \! \log_2(1 \! + \!\iota_k) \! - \! \iota_k \! + \! \frac{ (1+\iota_k) |\mathbf{w}_k^H \widetilde{\mathbf{H}}_{k,k-1} \mathbf{p}_k |^2}{\Gamma_k + \| \mathbf{w}_k \|^2_F \sigma_k^2}  \Bigg),
\end{aligned}
\end{equation}
where $\boldsymbol{\iota} \triangleq [\iota_1, \cdots, \iota_K]^\top \in \mathbb{R}^{K}$ is the auxiliary vector. $\Gamma_k$ is defined by
\begin{equation}
    \Gamma_k  = I_k (\mathbf{w}_k, \{  \mathbf{p}_i \}_{i \neq k, k+1}, \mathbf{\Phi}) + |\mathbf{w}_k^H \widetilde{\mathbf{H}}_{k,k-1} \mathbf{p}_k |^2.
\end{equation}

\bpara{Quadratic Transformation.} To further simplify the fractional term, we apply the quadratic transformation \cite{shen2018fractional1}, converting the components into integral expressions. The retransformed objective function is given by
\begin{equation}
\begin{aligned}
     & f_\tau(\mathbf{P}, \mathbf{W}, \mathbf{\Phi}, \boldsymbol{\iota}, \boldsymbol{\tau}) = \sum_{k \in \mathcal{K}} \alpha_k \Big( \log_2(1+\iota_k)\! -\! \iota_k \! \\
    & + \!{2\sqrt{1+\iota_k}  \Re{\tau_k^* \mathbf{w}_k^H \widetilde{\mathbf{H}}_{k,k-1} \mathbf{p}_k}} \! \\
    & - \! |\tau_k|^2 (\Gamma_k + \| \mathbf{w}_k \|^2_F \sigma^2) \! \Big),
\end{aligned}
\end{equation}
where $\boldsymbol{\tau} \triangleq [\tau_1, \cdots, \tau_K]^\top \in \mathbb{R}^{K}$ is another introduced auxiliary vector, leading to the reformulated optimization problem
\begin{maxi!}|s|[2]                   % mini! = minimize 
    {\substack{\mathbf{\Phi}, \mathbf{P}, \mathbf{W},\\ \boldsymbol{\iota}, \boldsymbol{\tau}}}                               % optimization variable
    {f_\tau(\mathbf{P}, \mathbf{W}, \mathbf{\Phi},\boldsymbol{\iota}, \boldsymbol{\tau}) \label{eq:op2}}   % objective function and label
    {\label{eq:p2}}             % label for opt problem
    {\mathcal{P}2:} 
    % optimization result
    {} 
    % \addConstraint{\|\mathbf{p}_k\|^2_F}{\leq P_{k-1}, \quad \forall k \in \mathcal{K} \label{eq:op2c1}}
    % \addConstraint{\|\mathbf{w}_k\|^2_F}{= 1, \quad \forall k \in \mathcal{K}\label{eq:op2c2}}
    % \addConstraint{\mathbf{\Phi}}{\in\mathcal{R}_i, \,  i \in\{1,2\}, \, \text{select one value for } i \label{eq:op2c4}}
    % \addConstraint{\mathbf{\Phi}}{\in \mathcal{S}_\ell, \, \ell \in \{ 1,2,3 \}, \, \text{select one value for } \ell \label{eq:op2c3}}.
    \addConstraint{\eqref{eq:op1c1}, \eqref{eq:op1c2}, \eqref{eq:op1c4}, \eqref{eq:op1c3}.}{\nonumber}
\end{maxi!}

\section{Solution to  Sum-rates Maximization}
\label{sec:algo}
The reformulated problem $\mathcal{P}2$ is a multi-variable optimization problem with non-convex constraints. To solve it, we employ the block coordinate descent (BCD) method \cite{bertsekas1997nonlinear}. In this framework, all variables are fixed except for the optimized variable in one iteration. This process is repeated until convergence. Regarding the constraints, there are three main difficulties: \textit{i}) constraints \eqref{eq:op1c4} and \eqref{eq:op1c3} are coupled, \textit{ii}) the symmetry constraint \eqref{eq:op1c4}, and \textit{iii}) the unitary constraint \eqref{eq:op1c3} complicates the problem. Therefore, to decouple \eqref{eq:op1c4} and \eqref{eq:op1c3}, we adopt the PDD method, which introduces a copy of the scattering matrix and penalty terms \cite{shi2020penalty}. To handle the symmetry constraint \eqref{eq:op1c4}, linearization is applied to the scattering matrix to ensure that only the required elements are optimized. With the first two difficulties addressed, the sub-problem with the unitary constraint \eqref{eq:op1c3} becomes an orthogonal Procrustes problem. This problem can be solved using the singular value decomposition (SVD) method and has a closed-form solution. The proposed design algorithm is summarized in Algorithm \ref{alg:alg1}.
\begin{algorithm}[t]
	\caption{Proposed Algorithm for  Sum-rate Design}
	\label{alg:alg1}
	\KwIn{$\mathbf{H}_\mathsf{SI}, \mathbf{H}_k, k \in \mathcal{K}$.}  
	\KwOut{$\mathbf{\Phi}^\mathsf{opt}, \mathbf{P}^\mathsf{opt}, \mathbf{W}^\mathsf{opt}$.} 
	\BlankLine
	Initialize $\mathbf{\Phi}, \mathbf{P}, \mathbf{W}, t=1$.
	
	\While{\textnormal{no convergence of objective function \eqref{eq:op2} \textbf{\&}} $\quad t<t_\mathsf{max}$ }{
        Update $\boldsymbol{\iota}_k^\mathsf{opt}, k \in \mathcal{K}$ by \eqref{eq:sinr}. \\
        Update $\boldsymbol{\tau}_k^\mathsf{opt}, k \in \mathcal{K}$ by \eqref{eq:tau}. \\
        Update $\mathbf{P}^\mathsf{opt}$ by \eqref{eq:poptimal}. \\
        Update $\mathbf{W}^\mathsf{opt}$ by \eqref{eq:woptimal}. \\
		Update $\mathbf{\Phi}^\mathsf{opt}$ by Algorithm \ref{alg:alg2}. \\
        $t = t + 1$.\\
	}	
	Return $\mathbf{\Phi}^\mathsf{opt}, \mathbf{P}^\mathsf{opt}, \mathbf{W}^\mathsf{opt}$.
\end{algorithm}
The details of the sub-problems are presented in the following subsections. Additionally, the optimal solutions for each block are derived. The optimal solution of auxiliary vectors $\boldsymbol{\iota}$ is derived in \cite{shen2018fractional1}, \ie $\iota_k^\mathsf{opt} = \gamma_{k}$.

% \bpara{Auxiliary Vectors: Block $\boldsymbol{\iota}$.} 
% Given fixed values for $\mathbf{P}, \mathbf{W}, \mathbf{\Phi},$ and $\boldsymbol{\tau}$, the sub-problem with respect to (w.r.t) $\boldsymbol{\iota}$ is convex and unconstrained. Therefore, the optimal solution can be derived by computing the derivative w.r.t $\boldsymbol{\iota}$ and setting it to zero, \ie $\frac{\partial f_\tau(\mathbf{P}, \mathbf{W}, \mathbf{\Phi},\boldsymbol{\iota}, \boldsymbol{\tau})}{\partial \boldsymbol{\iota}} = \mathbf{0}$. The optimal $\boldsymbol{\iota}^\mathsf{opt}_k, \forall k \in \mathcal{K}$ is:
% % \begin{equation}
% %     \iota_k^\mathsf{opt} = \gamma_{k} =\frac{ |\mathbf{w}_k^H \widetilde{\mathbf{H}}_{k, k-1} \mathbf{p}_k |^2}{ I_k (\mathbf{w}_k, \{ \mathbf{p}_i \}_{i \neq k, k+1}, \mathbf{\Phi}) + \| \mathbf{w}_k \|^2_F \sigma^2}, \, \forall k \in \mathcal{K}.
% % \label{eq:iota}
% % \end{equation}
% \begin{equation}
%     \iota_k^\mathsf{opt} = \gamma_{k} =\frac{ |\mathbf{w}_k^H \widetilde{\mathbf{H}}_{k, k-1} \mathbf{p}_k |^2}{ I_k (\mathbf{w}_k, \{ \mathbf{p}_i \}_{i \neq k, k+1}, \mathbf{\Phi}) + \| \mathbf{w}_k \|^2_F \sigma^2}.
% \label{eq:iota}
% \end{equation}

\bpara{Auxiliary Vectors: Block $\boldsymbol{\tau}$.} The sub-problem is an unconstrained convex optimization when $\mathbf{P}, \mathbf{W}, \mathbf{\Phi}, \boldsymbol{\iota}$ are fixed. Consequently, the optimal $\boldsymbol{\tau}^\mathsf{opt}$ is obtained by setting $\frac{\partial f_\tau(\mathbf{P}, \mathbf{W}, \mathbf{\Phi},\boldsymbol{\iota}, \boldsymbol{\tau})}{\partial \boldsymbol{\tau}} = \mathbf{0}$, which yields
\begin{equation}
\tau_k^\mathsf{opt} = \frac{\sqrt{1+\iota_k} \mathbf{w}_k^H \widetilde{\mathbf{H}}_{k,k-1} \mathbf{p}_k  }
{\Gamma_k  + \| \mathbf{w}_k \|^2_F \sigma^2}.
\label{eq:tau}
\end{equation}

\bpara{Transmit Precoder: Block $\mathbf{P}$.} Given that $\mathbf{W}, \mathbf{\Phi},\boldsymbol{\iota}$, and $\boldsymbol{\tau}$ are fixed, we isolate the terms associated with $\mathbf{P}$ given by
% \begin{equation}
% \begin{aligned}
%      f_\tau(\mathbf{P}) \! &  =  \sum_{k \in \mathcal{K}} \alpha_k  \Big( \! {2 \sqrt{1+\iota_k}  \Re{\tau_k^* \mathbf{w}_k^H \widetilde{\mathbf{H}}_{k-1} \mathbf{p}_k}}
%     \! \\
%     & \!-\! |\tau_k|^2 \! \big( | \mathbf{w}_k^H \mathbf{H}_{\mathsf{SI},k} \mathbf{p}_{k+1} + \mathbf{w}_k^H \widetilde{\mathbf{H}}_{k} \mathbf{p}_{k+1}|^2 \\
%     & + \! \mathbf{p}_k^H \widetilde{\mathbf{H}}_{k-1}^H \! \!  \sum_{p \in \mathcal{K}} \! \mathbf{w}_p \mathbf{w}_p^H \widetilde{\mathbf{H}}_{k-1} \mathbf{p}_k   \big)\Big).
%     \end{aligned}
% \end{equation}
\color{black}
\begin{equation}
\begin{aligned}
     f_\tau(\mathbf{P}) \! &  =  \sum_{k \in \mathcal{K}} \alpha_k  \Big( \! {2 \sqrt{1+\iota_k}  \Re{\tau_k^* \mathbf{w}_k^H \widetilde{\mathbf{H}}_{k, k-1} \mathbf{p}_k}}
    \! \\
    & \!-\! |\tau_k|^2 \! \big( | \mathbf{w}_k^H \mathbf{H}_{\mathsf{SI},k} \mathbf{p}_{k+1} + \mathbf{w}_k^H \overline{\mathbf{H}}_k \mathbf{p}_{k+1}|^2 \\
    & +  \sum_{i \in \mathcal{K}, i \neq k+1} \mathbf{p}_i^H \widetilde{\mathbf{H}}_{k,i-1}^H \mathbf{w}_k \mathbf{w}_k^H \widetilde{\mathbf{H}}_{k,i-1} \mathbf{p}_i  \big)\Big).
    \end{aligned}
\end{equation}
\color{black}
Thus, the optimization problem w.r.t $\mathbf{P}$ is expressed as follows:
\vspace{-10pt}
\begin{maxi!}|s|[2]                   % mini! = minimize 
    {\mathbf{P}}                               % optimization variable
    {f_\tau(\mathbf{P}) \label{eq:op3}}   % objective function and label
    {\label{eq:p3}}             % label for opt problem
    {\mathcal{P}3:} 
    % optimization result
    {} 
    \addConstraint{\|\mathbf{p}_k\|^2_F}{\leq P_{k-1}, \quad \forall k \in \mathcal{K} \label{eq:op3c1}}.
\end{maxi!}
Since both the objective function \eqref{eq:op3} and the constraint \eqref{eq:op3c1} are convex, the constraint can be incorporated into the objective function using the Lagrange multiplier method, based on the Karush–Kuhn–Tucker (KKT) conditions, by introducing a multiplier $\mu$. The components w.r.t $\mathbf{p}_k$ are
\begin{equation}
\begin{aligned}
     & f_\tau(\mathbf{p}_k)   =  \alpha_k  \Big( \! {2 \sqrt{1+\iota_k} \Re{\tau_k^* \mathbf{w}_k^H \widetilde{\mathbf{H}}_{k, k-1} \mathbf{p}_k}}
    \\
    & - |\tau_k|^2 \big(  \mathbf{p}_k^H \widetilde{\mathbf{H}}_{k,k-1}^H \! \!  \sum_{p \in \mathcal{K}} \! \mathbf{w}_p \mathbf{w}_p^H \widetilde{\mathbf{H}}_{k,k-1} \mathbf{p}_k   \big) \Big)\\
    & - \alpha_{k-1} |\tau_{k-1}|^2  | \mathbf{w}_{k-1}^H \mathbf{H}_{\mathsf{SI},k-1} \mathbf{p}_{k} \textcolor{black}{+ \mathbf{w}_{k-1}^H \overline{\mathbf{H}}_{k-1} \mathbf{p}_{k}|^2.}
    \end{aligned}
\end{equation}
The unconstrained sub-problem is given below
\begin{maxi!}|s|[2]                   % mini! = minimize 
    {\mathbf{p}_k}                               % optimization variable
    {f_\tau(\mathbf{p}_k) - \mu (\|\mathbf{p}_k\|^2_F - P_{k-1}). \label{eq:op4}}   % objective function and label
    {\label{eq:p4}}             % label for opt problem
    {\mathcal{P}4:} 
    % optimization result
    {} 
\end{maxi!}
The optimal solution for each precoder $\mathbf{p}_k$ can be derived using the first-order optimality condition:
\begin{equation}
\begin{aligned}
    \mathbf{p}_k^\mathsf{opt}   = &  \big( 
    \alpha_k |\tau_k|^2 \boldsymbol{\zeta}_{1,k} \! + \alpha_{k-1} \! |\tau_{k-1}|^2 \boldsymbol{\zeta}_{2,k}  + \mu^\mathsf{opt} \mathbf{I}  \big)^{-1} \\
    &\alpha_k \sqrt{ 1+\iota_k} \tau_k \widetilde{\mathbf{H}}_{k,k-1}^H \mathbf{w}_k,
\label{eq:poptimal}
\end{aligned}
\end{equation}
where $\boldsymbol{\zeta}_{1,k}$ and $\boldsymbol{\zeta}_{2,k}$ are defined by
\begin{equation}
    \boldsymbol{\zeta}_{1,k} =  \widetilde{\mathbf{H}}_{k,k-1}^H \! \!  \sum_{p \in \mathcal{K}} \! \mathbf{w}_p \mathbf{w}_p^H \widetilde{\mathbf{H}}_{k,k-1},
\end{equation}
% \begin{equation}
% \begin{aligned}
%     \boldsymbol{\zeta}_{2,k}  = &  \big( \mathbf{H}_{\mathsf{SI},k-1}^H \mathbf{w}_{k-1} \mathbf{w}_{k-1}^H \mathbf{H}_{\mathsf{SI},k-1} \\
%     & +  \widetilde{\mathbf{H}}_{k-1,k-1}^H \mathbf{w}_{k-1} \mathbf{w}_{k-1}^H \widetilde{\mathbf{H}}_{k-1,k-1}\\
%     & + 2\Re{\mathbf{H}_{\mathsf{SI},k-1}^H \mathbf{w}_{k-1} \mathbf{w}_{k-1}^H \widetilde{\mathbf{H}}_{k-1,k-1}} \big),
% \end{aligned}
% \end{equation}
\color{black}
\begin{equation}
    \begin{aligned}
        \boldsymbol{\zeta}_{2,k}  = &  \big( \mathbf{H}_{\mathsf{SI},k-1}^H \mathbf{w}_{k-1} \mathbf{w}_{k-1}^H \mathbf{H}_{\mathsf{SI},k-1} +  \overline{\mathbf{H}}_{k-1}^H \mathbf{w}_{k-1} \mathbf{w}_{k-1}^H \overline{\mathbf{H}}_{k-1})\\
        & + 2\Re{\mathbf{H}_{\mathsf{SI},k-1}^H \mathbf{w}_{k-1} \mathbf{w}_{k-1}^H \overline{\mathbf{H}}_{k-1}^H} \big),
    \end{aligned}
\end{equation}
\color{black}
and we can use bisection search to find $\mu^\mathsf{opt}$.

\bpara{Receive Combiner: Block $\mathbf{W}$.}
Given fixed values for $\mathbf{P}, \mathbf{\Phi},\boldsymbol{\iota}$, and $\boldsymbol{\tau}$, the objective function w.r.t $\mathbf{W}$ is expressed as
\color{black}
\begin{equation}
\begin{aligned}
     &f_\tau(\mathbf{W}) \!  =  \sum_{k \in \mathcal{K}} \alpha_k  \Bigg( \! {2 \sqrt{1+\iota_k}  \Re{\tau_k^* \mathbf{w}_k^H \widetilde{\mathbf{H}}_{k,k-1} \mathbf{p}_k}}
    \! \\
    & \!-\! |\tau_k|^2  \! \bigg( | \mathbf{w}_k^H \mathbf{H}_{\mathsf{SI},k} \mathbf{p}_{k+1} 
     + \mathbf{w}_k^H \overline{\mathbf{H}}_k \mathbf{p}_{k+1}|^2\\
    & + \! \mathbf{w}_k^H ( \! \! \! \! \! \! \sum_{i \in \mathcal{K}, i \neq k+1} \! \! \!\! \! \widetilde{\mathbf{H}}_{k,i-1} \mathbf{p}_i \mathbf{p}_i^H \widetilde{\mathbf{H}}_{k,i-1}^H
    ) \mathbf{w}_k  + \| \mathbf{w}_k \|^2_F \sigma^2   \bigg) \! \!\Bigg).
    \end{aligned}
\end{equation}
\color{black}
Therefore, the sub-optimization problem with regard to $\mathbf{W}$ is 
\vspace{-10pt}
\begin{maxi!}|s|[2]                   % mini! = minimize 
    {\mathbf{W}}                               % optimization variable
    {f_\tau(\mathbf{W}) \label{eq:op5}}   % objective function and label
    {\label{eq:p5}}             % label for opt problem
    {\mathcal{P}5:} 
    % optimization result
    {}
    \addConstraint{\|\mathbf{w}_k\|^2_F}{= 1, \quad \forall k \in \mathcal{K}. \label{eq:op4c1}}
\end{maxi!}

The constraint \eqref{eq:op4c1} is non-convex, making the problem hard to tackle with. Therefore, we relax this constraint and the sub-problem becomes unconstrained. Finally, the solution is normalized to satisfy normalization constraint. Specifically,
% first treat the sub-problem as an unconstrained optimization problem w.r.t $\mathbf{W}$. 
We still take the derivative of the objective function w.r.t $\mathbf{w}_k$ and set it to zero, leading to the optimal solution for each combiner $\mathbf{w}_k$. 
% Once the algorithm converges, we normalize the obtained solution to satisfy the unitary constraint \eqref{eq:op4c1}. 
The expression of the optimal combiner is
\begin{equation}
     \mathbf{w}_k^\mathsf{opt} \! = \! \big(  |\tau_k|^2 \boldsymbol{\xi}_k \big)^{-1}
    (\sqrt{1+\iota_k}  \tau_k^* \widetilde{\mathbf{H}}_{k,k-1} \mathbf{p}_k),
\label{eq:woptimal}
\end{equation}
where $\boldsymbol{\xi}$ is given by
\color{black}
\begin{equation}
\begin{aligned}
   & \boldsymbol{\xi}_k  =  \big( \mathbf{H}_{\mathsf{SI},k} \mathbf{p}_{k+1} \mathbf{p}_{k+1}^H \mathbf{H}^H_{\mathsf{SI},k} + 2 \Re{\overline{\mathbf{H}}_{k} \mathbf{p}_{k+1} \mathbf{p}_{k+1}^H \mathbf{H}^H_{\mathsf{SI},k}} \\
    & + \overline{\mathbf{H}}_{k}  \mathbf{p}_{k+1} \mathbf{p}_{k+1}^H  \overline{\mathbf{H}}_{k}^H
    \big)  + \! \! \! \! \sum_{i \in \mathcal{K}, i \neq k+1} \! \! \! \widetilde{\mathbf{H}}_{k,i-1} \mathbf{p}_i \mathbf{p}_i^H \widetilde{\mathbf{H}}_{k,i-1}^H + \sigma^2.
    \end{aligned}
\end{equation}
\color{black}
Finally, we normalize the obtained solution $\mathbf{w}_k^\mathsf{opt}$ to satisfy the constraint \eqref{eq:op4c1}.

\bpara{BD-RIS Scattering Matrix: $\mathbf{\Phi}$.} Given fixed values for $\mathbf{P}, \mathbf{W}, \boldsymbol{\iota}, \boldsymbol{\tau}$, we isolate the terms related to $\mathbf{\Phi}$, expressed as
\color{black}
\begin{equation}
\begin{aligned}
     & f_\tau(\mathbf{\Phi})  = \sum_{k \in \mathcal{K}} \alpha_k \Big( {2\sqrt{1+\iota_k } \Re{\tau_k^* \mathbf{w}_k^H \widetilde{\mathbf{H}}_{k,k-1} \mathbf{p}_k}} \! \\
     & \qquad - |\tau_k|^2 \big(  | \mathbf{w}_k^H \mathbf{H}_{\mathsf{SI},k} \mathbf{p}_{k+1} 
     \! + \! \mathbf{w}_k^H \overline{\mathbf{H}}_{k} \mathbf{p}_{k+1}|^2 \! \\
    & \qquad + \sum_{i \in \mathcal{K}, i \neq k+1}  \! | \mathbf{w}_k^H \widetilde{\mathbf{H}}_{k,i-1} \mathbf{p}_{i}|^2
     \big) \! \Big)\\
     & = 2 \Re{\Tr(\mathbf{C}_1 \mathbf{\Phi})} - 2 \Re{\Tr(\mathbf{C}_2 \mathbf{\Phi})}   - \Tr(\mathbf{A} \mathbf{\Phi} \mathbf{B} \mathbf{\Phi}^H) \\
     & + 2 \Re{\Tr( \mathbf{C}_3 \mathbf{\Phi})} - 2 \Re{\Tr( \mathbf{C}_4 \mathbf{\Phi})} + 2 \Re{\Tr( \mathbf{C}_5 \mathbf{\Phi})}.
\end{aligned}
\end{equation}
\color{black}
Table \ref{tab:1} provides the definitions of the newly introduced notations. Since the single- and fully-connected BD-RIS are special cases of the group-connected case, the following sub-problem w.r.t. the group-connected BD-RIS is used for explanation:
\begin{maxi!}|s|[2]                   % mini! = minimize 
    {\mathbf{\Phi}}                               % optimization variable
    {f_\tau(\mathbf{\Phi}) \label{eq:op6}}   % objective function and label
    {\label{eq:p6}}             % label for opt problem
    {\mathcal{P}6:} 
    % optimization result
    {}
    \addConstraint{\mathbf{\Phi}\in\mathcal{R}_i, \,  i \in\{1,2\},}{\label{eq:op6c2}}
    \addConstraint{\mathbf{\Phi}}{\in \mathcal{S}_1 \label{eq:op6c1}.}
\end{maxi!}
The main challenges in solving the sub-problem $\mathcal{P}6$ are the unitary constraint \eqref{eq:op6c1} and the symmetry constraint \eqref{eq:op6c2}. Below, we present a general approach for designing $\mathbf{\Phi}$ applicable to both NR-BD-RIS and R-BD-RIS.

\begin{algorithm}[t]
	\caption{Algorithm for Optimizing BD-RIS Scattering Matrix $\mathbf{\Phi}$}
	\label{alg:alg2}
	\KwIn{$\mathbf{P}, \mathbf{W}, \boldsymbol{\iota}_k, \boldsymbol{\tau}_k,  \mathbf{H}_\mathsf{SI}, \mathbf{H}_k, k \in \mathcal{K}$.}  
	\KwOut{$\mathbf{\Phi}^\mathsf{opt}$.} 
	% \BlankLine
	Initialize $\{\mathbf{\Phi}_g\}, \{\mathbf{\Psi}_g\}, \{\mathbf{\Lambda}_g\}, \rho, t_\mathsf{inner}=t_\mathsf{outer}=1$.
 
    \For{$g \gets 1$ \KwTo $G$}{
        \While{\textnormal{$\| \mathbf{\Phi}_g - \mathbf{\Psi}_g \|_\infty > \varepsilon$ \textbf{\&}} $ t_\mathsf{outer}<t_\mathsf{outer\, max}$ }{
        \While{\textnormal{no convergence of objective function \eqref{eq:op7} \textbf{\&}} $ t_\mathsf{inner}<t_\mathsf{inner\, max}$ }{
        Update $ \mathbf{\Phi}_g $ by \eqref{eq:phi}.\\
        Update $ \mathbf{\Psi}_g $ by \eqref{eq:phi_copy}.\\
        $t_\mathsf{inner} = t_\mathsf{inner} + 1$.
	}
    \uIf{$\| \mathbf{\Phi}_g - \mathbf{\Psi}_g \|_\infty < \epsilon$}{
        $\mathbf{\Lambda}_g = \mathbf{\Lambda}_g + \rho^{-1} (\mathbf{\Phi}_g - \mathbf{\Psi}_g)$.
    }
    \Else{
        $\rho = c\rho$.
    }
    $t_\mathsf{outer} = t_\mathsf{outer} + 1$.
	}	
    }
	\KwRet{ $\mathbf{\Phi}^\mathsf{opt} = \operatorname{blkdiag}(\mathbf{\Phi}^\mathsf{opt}_1, \cdots, \mathbf{\Phi}^\mathsf{opt}_G)$.}
\end{algorithm}

\begin{table}[t]
    \color{black}
    \caption{Newly Introduced Notations}
    \resizebox{0.48\textwidth}{!}{
    \renewcommand{\arraystretch}{1.8}
    \begin{tabular}{|l|l|}
    \hline
    $\mathbf{A} = \sum_{k \in \mathcal{K}} \alpha_k |\tau_k|^2 \mathbf{H}_{\mathsf{ref},k}^* \mathbf{w}_{k} \mathbf{w}_{k}^H \mathbf{H}_{\mathsf{ref},k}^\top$ &
    $\mathbf{B} = \sum_{i \in \mathcal{K}} \mathbf{H}_{\mathsf{ref},i-1} \mathbf{p}_{i} \mathbf{p}_{i}^H  \mathbf{H}_{\mathsf{ref},i-1}^H $ \\ \hline
    $\mathbf{C}_1 = \sum_{k \in \mathcal{K}} \alpha_k \sqrt{1+\iota_k} \tau_k^* \mathbf{H}_{\mathsf{ref},k-1} \mathbf{p}_k \mathbf{w}_k^H \mathbf{H}_{\mathsf{ref},k}^\top  $ & $\mathbf{C}_3 = \mathbf{A} \mathbf{B}$ \\ \hline
    \multicolumn{2}{|l|}{$\mathbf{C}_2 = \sum_{k \in \mathcal{K}} \alpha_k |\tau_k|^2  \mathbf{H}_{\mathsf{ref},k} \mathbf{p}_{k+1} \mathbf{p}_{k+1}^H \mathbf{H}_{\mathsf{SI},k}^H \mathbf{w}_k \mathbf{w}_k^H \mathbf{H}_{\mathsf{ref},k}^\top$} \\ \hline
    \multicolumn{2}{|l|}{$\mathbf{C}_4 = \sum_{k \in \mathcal{K}} \alpha_k |\tau_k|^2  \sum_{i \in \mathcal{K}} \mathbf{H}_{\mathsf{ref}, i-1} \mathbf{p}_{i} \mathbf{p}_{i}^H \mathbf{H}_{\mathsf{dir},k,i-1}^H \mathbf{w}_k \mathbf{w}_k^H \mathbf{H}_{\mathsf{ref},k}^\top$} \\ \hline
    \multicolumn{2}{|l|}{$\mathbf{C}_5 = \sum_{k \in \mathcal{K}} \alpha_k |\tau_k|^2  \mathbf{H}_{\mathsf{ref},k} \mathbf{p}_{k+1} \mathbf{p}_{k+1}^H \mathbf{H}_{\mathsf{dir},k,k}^H \mathbf{w}_k \mathbf{w}_k^H \mathbf{H}_{\mathsf{ref},k}^\top$} \\ \hline
    \end{tabular}
    \label{tab:1}
    }
    \color{black}
    \vspace{-10pt}
\end{table}

\subsection{PDD Method to Decouple \eqref{eq:op6c1} and \eqref{eq:op6c2}}
In this section, we employ the PDD method \cite{shi2020penalty, zhou2023optimizing} to decouple constraints \eqref{eq:op6c1} and \eqref{eq:op6c2}. The PDD framework is a two-loop iterative approach \cite{shi2020penalty}, comprising an inner loop and an outer loop. The inner loop addresses the augmented Lagrangian problem using BCD approach, and the outer loop adjusts the Lagrangian dual variables $\{ \mathbf{\Lambda}_g \}$ and the penalty coefficient $\rho$ until convergence (\ie $\| \mathbf{\Phi} - \mathbf{\Psi} \|_\infty \leq \varepsilon$). Specifically, in the inner loop, each $\mathbf{\Phi}_g, \forall g \in \mathcal{G}$, is optimized iteratively. The outer loop adjusts $\{ \mathbf{\Lambda}_g \}$ and $\rho$ to ensure the algorithm converges. This algorithm is concluded in Algorithm \ref{alg:alg2}. To decouple constraints \eqref{eq:op6c1} and \eqref{eq:op6c2}, we introduce a copy $\{\mathbf{\Psi}_g\}$ of $\{\mathbf{\Phi}_g\}$ and enforce the equality constraint $\mathbf{\Psi}_g = \mathbf{\Phi}_g, \forall g \in \mathcal{G}$. We then have the following problem:
\vspace{-5pt}
\begin{maxi!}|s|[2]                   % mini! = minimize 
    {\mathbf{\Phi}, \mathbf{\Psi}}                               % optimization variable
    {f_\tau(\mathbf{\Phi}) \label{eq:op70}}   % objective function and label
    {\label{eq:p70}}             % label for opt problem
    {\mathcal{P}7:} 
    % optimization result
    {}
    \addConstraint{\mathbf{\Psi}_g^H \mathbf{\Psi}_g = \mathbf{I},}{\quad  \forall g \in \mathcal{G} \label{eq:op70c1},}
    \addConstraint{\mathbf{\Phi}_g = \mathbf{\Psi}_g,}{\quad  \forall g \in \mathcal{G} \label{eq:op70c2},}
    \addConstraint{\eqref{eq:op6c2}.}{\nonumber}
\end{maxi!}
Subsequently, we reformulate the problem by introducing a penalty term to enforce the equality constraint \eqref{eq:op70c2} with the Lagrangian dual variable and penalty coefficient $\mathbf{\Lambda}_g \in \mathbb{C}^{M \times M}, \forall g \in \mathcal{G}$, and $1/\rho$, respectively. This allows us to optimize $\mathbf{\Phi}_g$ and $\mathbf{\Psi}_g$ separately while ensuring they converge to the same solution. The reformulated problem is given by
\begin{equation}
\begin{aligned}
\label{eq:obj_pdd}
    L( \{ \!\mathbf{\Phi}_g \!\}, \{\!\mathbf{\Psi}_g\! \}, &\{\!\mathbf{\Lambda}_g \!\},\rho) \! =\! - \!f_\tau(\mathbf{\Phi}) \!+  \!\frac{1}{2 \rho}  \!\sum_{g \in \mathcal{G}}\! \|\mathbf{\Phi}_g \! - \! \mathbf{\Psi}_g \|^2 \\
    & + \sum_{g \in \mathcal{G}} \Re{\Tr(\mathbf{\Lambda}_g^H(\mathbf{\Phi}_g - \mathbf{\Psi}_g))}, 
\end{aligned}
\end{equation}
We then have the reformulated problem as 
\begin{mini!}|s|[2]                   % mini! = minimize 
    {\mathbf{\Phi}, \mathbf{\Psi}}                               % optimization variable
    {L( \{ \mathbf{\Phi}_g \}, \{\mathbf{\Psi}_g \}, \{\mathbf{\Lambda}_g \},\rho\,) \label{eq:op7}}   % objective function and label
    {\label{eq:p7}}             % label for opt problem
    {\mathcal{P}8:} 
    % optimization result
    {}
    \addConstraint{\eqref{eq:op6c2}, \eqref{eq:op70c1}.}{\nonumber}
\end{mini!}
% The inner and outer loop update processes are provided below.
\vspace{-10pt}
\subsection{Inner Loop}
\begin{figure}[t]
    \centering
    \includegraphics[width=0.38\textwidth]{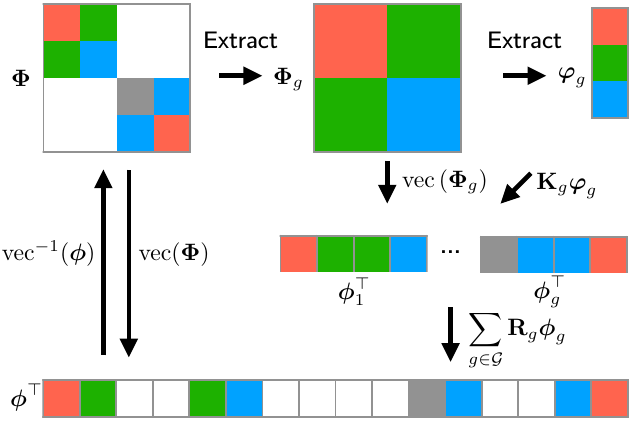}
    \caption{The linear reformulation and reconstruction for the reciprocal case.}
    \label{fig:linear}
\end{figure}

In the inner loop, we first preprocess the scattering matrix $\mathbf{\Phi}_g$ to a linear formulation, which is explained in \fig{fig:linear}. Subsequently, we update $\mathbf{\Phi}_g$ and $\mathbf{\Psi}_g$.

\subsubsection{Linear Reformulation of Symmetry Constraint}
We utilize the linear reformulation to tackle the symmetry constraint \eqref{eq:op70c1}. We extract the necessary elements of the scattering matrix $\mathbf{\Phi}_g$ and store them in a vector $\boldsymbol{\varphi}_g$. For NR-BD-RIS, all elements of $\mathbf{\Phi}_g$ are extracted, while for R-BD-RIS, only the diagonal and lower-triangular elements are extracted. The vector $\boldsymbol{\varphi}_g$ is constructed in a column-wise order. The reconstruction of the full vector $\boldsymbol{\phi}_g = \operatorname{vec}(\boldsymbol{\Phi}_g)$ from $\boldsymbol{\varphi}_g$ is achieved using a permutation matrix $\mathbf{K}_{g}$ defined in \cite{liu2024non}, which depends on whether it is NR-BD-RIS or R-BD-RIS.
% To address the symmetry constraint, we focus on optimizing the necessary elements of the matrix: $i)$ all elements of $\mathbf{\Phi}_g$ for NR-BD-RIS, and $ii)$ the diagonal and lower-triangular elements of $\mathbf{\Phi}_g$ for R-BD-RIS. These elements are extracted and stored in a vector $\boldsymbol{\varphi}_g$ in a column-wise order. To reconstruct $\boldsymbol{\phi}_g = \operatorname{vec}\left(\boldsymbol{\Phi}_g\right)$ from $\boldsymbol{\varphi}_g$, a permutation matrix is introduced. The reconstruction process is expressed as
\begin{equation}
    \boldsymbol{\phi}_g \triangleq \operatorname{vec}\left(\boldsymbol{\Phi}_g\right)=\mathbf{K}_{g} \boldsymbol{\varphi}_g, \quad \forall g \in \mathcal{G}.
\label{eq:mat2vec}
\end{equation}
To position the group vector $\boldsymbol{\phi}_g$ into the overall vector $\boldsymbol{\phi}$, a reshaping matrix $\mathbf{R}_g \in \mathbb{B}^{M^2 \times M_g^2}$ is required and defined in \cite{liu2024non}. 
The whole mapping process is illustrated in \fig{fig:linear} and the mathmetical formulation is given by
\begin{equation}
\boldsymbol{\phi} \triangleq \operatorname{vec}(\boldsymbol{\Phi})=\sum_{g \in \mathcal{G}} \mathbf{R}_g \boldsymbol{\phi}_g,
\end{equation}
After the symmtery constraint is reformulated, we can apply the PDD method to optimize the scattering matrix $\mathbf{\Phi}$.

% \begin{equation}
% \boldsymbol{\phi} \triangleq \operatorname{vec}(\boldsymbol{\Phi})=\sum_{g \in \mathcal{G}} \mathbf{R}_g \boldsymbol{\phi}_g,
% \end{equation}
% where $\mathbf{R}_g, \forall g \in \mathcal{G}$ is defined as
% \begin{equation}
%     \mathbf{R}_g(i,j) = \left \{ 
%     \begin{aligned}
%         \begin{aligned}
%             1&, 
%             \begin{aligned}
%                 \text{if} \,\, i &= (M_g(g-1)+m-1)M \\
%                                  & \quad + (M_g(g-1)+n), \, \text{and}\\
%                                j &= (m-1)M_g + n\\
%             \end{aligned}\\
%             0&, \, \text{otherwise},
%         \end{aligned}
%     \end{aligned}
%      \right.
% \end{equation}
% where $1 \leq m \leq M_g$ and $1 \leq n \leq M_g$ are iteration indices. After preprocessing the constraints, the PDD method can be applied to optimize the scattering matrix.

\subsubsection{$\mathbf{\Phi}_g$ in the inner loop}
With following transformations with trace operation
\begin{equation}
\operatorname{Tr}(\mathbf{C}\boldsymbol{\Phi})\!=\!\operatorname{vec}^\top\left(\mathbf{C}^\top\right) \boldsymbol{\phi}\!=\!\sum_{g \in \mathcal{G}} \operatorname{vec}^\top\left(\mathbf{C}^\top\right) \mathbf{R}_g \mathbf{K}_{g} \boldsymbol{\varphi}_g,
\end{equation}
\vspace{-10pt}
\begin{equation}
\begin{aligned}
    \operatorname{Tr}&\left(\mathbf{B} \boldsymbol{\Phi} \mathbf{A} \boldsymbol{\Phi}^H\right)\!=\!\boldsymbol{\phi}^H\left(\mathbf{A}^\top \otimes \mathbf{B}\right) \boldsymbol{\phi} \\
    &=\big(\sum_{g \in \mathcal{G}} \mathbf{R}_g \mathbf{K}_{g} \boldsymbol{\varphi}_g\big)^H   \!\! \left(\mathbf{A}^\top \! \! \otimes \! \mathbf{B}\right) \!\big(\sum_{g \in \mathcal{G}} \mathbf{R}_g \mathbf{K}_{g} \boldsymbol{\varphi}_g \big),
\end{aligned}
\end{equation}
the sub-problem w.r.t $\mathbf{\varphi}_g$ is formulated as 
\begin{mini!}|s|[2]                   % mini! = minimize 
    {\boldsymbol{\varphi}_g}                               % optimization variable
    {L( \boldsymbol{\varphi}_g ) = \boldsymbol{\varphi}_g^H \boldsymbol{\Delta} \boldsymbol{\varphi}_g - 2 \Re{\boldsymbol{\varphi}_g^H \boldsymbol{\delta}}. \label{eq:op8}}   % objective function and label
    {\label{eq:p8}}             % label for opt problem
    {\mathcal{P}9:} 
    % optimization result
    {}
\end{mini!}
The vectorization is defined as $\boldsymbol{\psi}_g \triangleq \operatorname{vec}(\boldsymbol{\Psi}_g)$, $\lambda_g \triangleq \operatorname{vec}(\boldsymbol{\Lambda}_g)$,
% \begin{equation}
% \begin{aligned}
%      \boldsymbol{\Delta}  =  \mathbf{K}_{g}^H \mathbf{R}_g^H \big(
%     \mathbf{A}_1^\top \otimes \mathbf{I} + \mathbf{A}_2^\top \otimes \mathbf{B} - \mathbf{A}_3^\top \otimes \mathbf{I} \big) \mathbf{R}_g \mathbf{K}_{g} + \frac{1}{2 \rho} \mathbf{K}_{g}^H \mathbf{K}_{g},
% \end{aligned}
% \end{equation}
\begin{equation}
\begin{aligned}
     \boldsymbol{\Delta}  =  \mathbf{K}_{g}^H \mathbf{R}_g^H \big(
      \mathbf{A}^\top \otimes \mathbf{B}  \big) \mathbf{R}_g \mathbf{K}_{g} + \frac{1}{2 \rho} \mathbf{K}_{g}^H \mathbf{K}_{g},
\end{aligned}
\end{equation}
% \begin{equation}
% \begin{aligned}
%      \boldsymbol{\delta} = &  \mathbf{K}_{g}^H \mathbf{R}_g^H \bigg(
%     \operatorname{vec}^*(\mathbf{C}_1^\top) - \operatorname{vec}^*(\mathbf{C}_2^\top) + \operatorname{vec}^*(\mathbf{A}_1^\top) \\
%     & + \operatorname{vec}^*(\mathbf{C}_3^\top) - \operatorname{vec}^*(\mathbf{A}_3^\top)
%     \big) + \frac{1}{2\rho} \mathbf{K}_{g}^H \mathbf{K}_{g} \boldsymbol{\psi}_g - \frac{1}{2} \mathbf{K}_{g}^H \boldsymbol{\lambda}_g.
% \end{aligned}
% \end{equation}
\color{black}
\vspace{-10pt}
\begin{equation}
\begin{aligned}
    & \boldsymbol{\delta} =  \mathbf{K}_{g}^H \mathbf{R}_g^H \bigg(
    \operatorname{vec}^*(\mathbf{C}_1^\top) - \operatorname{vec}^*(\mathbf{C}_2^\top) + \operatorname{vec}^*(\mathbf{C}_3^\top) \\
    & -\! \operatorname{vec}^*(\mathbf{C}_4^\top)\! + \!\operatorname{vec}^*(\mathbf{C}_5^\top) \!
    \bigg) \!+\! \frac{1}{2\rho} \mathbf{K}_{g}^H \mathbf{K}_{g} \boldsymbol{\psi}_g\! - \!\frac{1}{2} \mathbf{K}_{g}^H \boldsymbol{\lambda}_g.
\end{aligned}
\end{equation}
\color{black}
$\mathcal{P}9$ is an unconstrained convex optimization problem, and the optimal solution can be obtained by solving $\frac{\partial{L( \boldsymbol{\varphi}_g ) }}{\partial{\boldsymbol{\varphi}_g}} = \mathbf{0}$. The optimal solution is then given by
\begin{equation}
    \boldsymbol{\varphi}_g^\mathsf{opt} = \boldsymbol{\Delta}^{-1} \boldsymbol{\delta}.
    \label{eq:phi}
\end{equation}

\subsubsection{$\mathbf{\Psi}_g$ in the inner loop} 
We fix the other variables and optimize $\mathbf{\Psi}_g$ in the inner loop. The optimization problem is formulated as follows:
% With fixed other variables, the optimization problem with regard to $\mathbf{\Psi}_g$ is given by
\begin{mini!}|s|[2]                   % mini! = minimize 
    {\boldsymbol{\Psi}_g}                               % optimization variable
    {\| \boldsymbol{\Psi}_g - (\rho \boldsymbol{\Lambda}_g + \boldsymbol{\Phi_g})  \|^2_F, \label{eq:op9}}   % objective function and label
    {\label{eq:p9}}             % label for opt problem
    {\mathcal{P}10:} 
    % optimization result
    {}
    \addConstraint{\mathbf{\Psi}_g^H \mathbf{\Psi}_g = \mathbf{I}.}{\label{eq:op9c1}}
\end{mini!}
This problem is called orthogonal procrustes problem and has a close-form solution \cite{gibson1962least, manton2002optimization, gloub1996matrix, zhou2023optimizing}. The solution is given by
% This is actually the orthogonal procrustes problem with a close-form solution \cite{gibson1962least, manton2002optimization, gloub1996matrix, zhou2023optimizing}, expressed as 
\begin{equation}
\boldsymbol{\Psi}_g^{\mathsf{opt}}=\mathbf{U}_g  \mathbf{V}_g^H.
\label{eq:phi_copy}
\end{equation}
Subsequently, we utilize singular vector decomposition (SVD) of $\rho \boldsymbol{\Lambda}_g + \boldsymbol{\Phi_g}$ to obtain unitary matrices $\mathbf{U}_g$, and $\mathbf{V}_g$.

\vspace{-10pt}
\subsection{Outer loop} 
After the inner loop converges, the dual variable $\mathbf{\Lambda}_g$ and the penalty coefficient $\rho$ are updated. The convergence condition is defined as $\| \boldsymbol{\Phi}_g - \boldsymbol{\Psi}_g \|_\infty < \epsilon$, where $\epsilon$ is a small positive constant. 
If the convergence condition is satisfied, we update the dual variable as follows:
\begin{equation}
    \mathbf{\Lambda}_g = \mathbf{\Lambda}_g + \rho^{-1} (\mathbf{\Phi}_g - \mathbf{\Psi}_g).
\end{equation}
Otherwise, if the convergence condition is not satisfied, this means that the equality between $\mathbf{\Phi}_g$ and $\mathbf{\Psi}_g$ is not reached. We thus update the penalty coefficient $\rho$ as $ \rho = c\rho$,
% Otherwise, if $\| \mathbf{\Phi}_g - \mathbf{\Psi}_g \|_\infty > \epsilon$, indicating a significant difference between $\mathbf{\Phi}_g$ and $\mathbf{\Psi}_g$, the penalty coefficient $\rho$ is adjusted to enforce equality:
% \begin{equation}
%     \rho = c\rho,
% \end{equation}
where $c \in (0,1)$ is the penalty scaling factor. Once the optimal $\mathbf{\Phi}_g^\mathsf{opt}$ for each group is obtained, the overall scattering matrix $\mathbf{\Phi}^\mathsf{opt}$ is reconstructed by combining the group matrices, \ie $\mathbf{\Phi}^\mathsf{opt} = \operatorname{blkdiag}(\mathbf{\Phi}_1, \cdots, \mathbf{\Phi}_G)$.

\vspace{-10pt}

\subsection{Initialization}
An appropriate initialization is needed for the scattering matrix $\boldsymbol{\Phi}$, all precoders $\mathbf{P}$ and all combiners $\mathbf{W}$ at the FD UEs. Specifically, we initialize $\boldsymbol{\Phi}$ as a diagonal matrix,
$
\boldsymbol{\Phi}=\operatorname{diag}(\phi_{1},\ldots,\phi_{M}),
$
where each diagonal entry has amplitude $1/\sqrt{2}$ and a random phase in $[0,2\pi)$.

At the $i^\mathrm{th}$ FD UE, with this initial BD-RIS scattering matrix, we adopt minimum mean square error (MMSE) precoder to initialize precoder for  $k^\mathrm{th}$ FD UE
\begin{equation}
\mathbf{p}_k
=\widetilde{\mathbf{H}}_{k,i}^H
\left(
\widetilde{\mathbf{H}}_{k,i}\widetilde{\mathbf{H}}_{k,i}^H
+\sigma^2\mathbf{I}
\right)^{-1}.
\end{equation}
We then normalize it to meet the transmit power constraint,
$
\mathbf{p}_k
=\frac{\sqrt{P_{k-1}}\mathbf{p}_k}{\|\mathbf{p}_k\|_F}.
$

At the $k^\mathrm{th}$ FD UE, using the same initial BD\mbox{-}RIS coefficients and the above precoder, the MMSE combiner is
\begin{equation}
    \mathbf{w}_k
=\left(
\widetilde{\mathbf{H}}_{k,i}\mathbf{p}_k\mathbf{p}_k^H\widetilde{\mathbf{H}}_{k,i}^H
+\sigma^2\mathbf{I}
\right)^{-1}
\widetilde{\mathbf{H}}_{k,i}\mathbf{p}_k,
\end{equation}
followed by normalization,
$
\mathbf{w}_k=\frac{\mathbf{w}_k}{\|\mathbf{w}_k\|_F}.
$

\color{black}

\subsection{Convergence Analysis}
\begin{figure}[t]
    \centering
    \includegraphics[width = 0.4\textwidth]{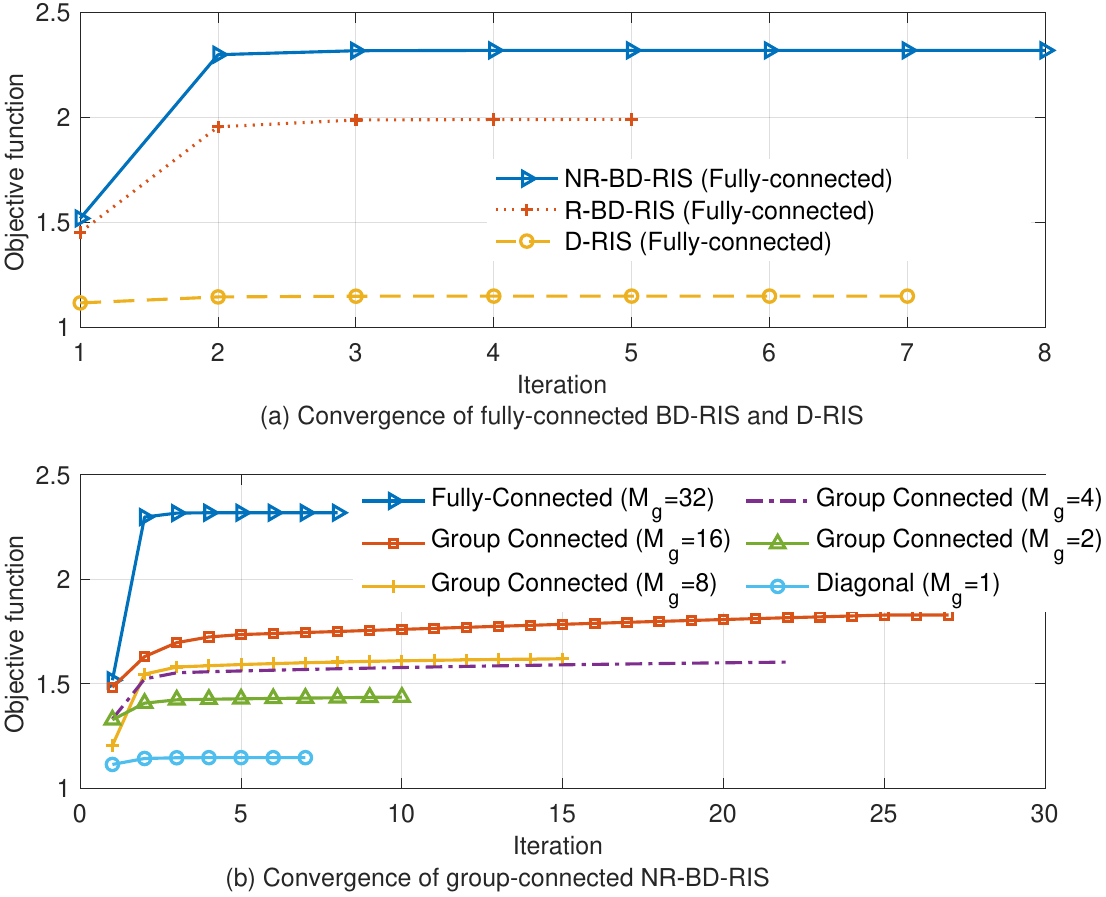}
    \centering
    \caption{Convergence analysis for Algorithm \ref{alg:alg1} with both direct and reflected links, using RIS elements $M=32$. The 3 FD UEs are located at $30^\circ$, $90^\circ$, and $150^\circ$, respectively.  (a) Convergence for fully-connected NR-BD-RIS, R-BD-RIS, and D-RIS. (b) Convergence for group-connected NR-BD-RIS with varying group sizes.}
    \label{fig:convergence}
\end{figure}
\begin{figure}[t]
    \centering
    \includegraphics[width = 0.4\textwidth]{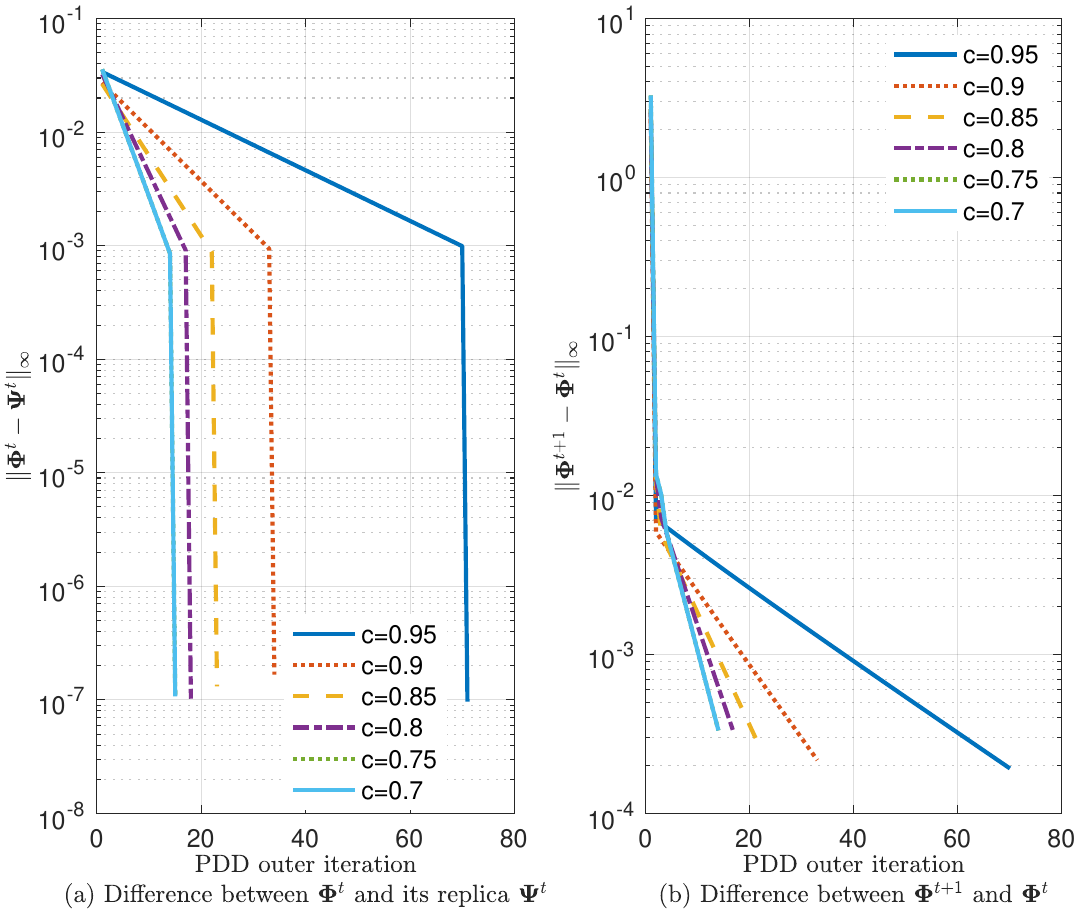}
    \centering
    \caption{Convergence analysis for Algorithm \ref{alg:alg2} (PDD) with both direct and reflected links for different $c$ values, using $M=32$ and $N = 1$. The 3 FD UEs are located at $30^\circ$, $90^\circ$, and $150^\circ$. (a) Convergence of the scattering matrix $\mathbf{\Phi}$ and its replica $\mathbf{\Psi}$, measured by $\| \mathbf{\Phi} - \mathbf{\Psi} \|_\infty$. (b) Convergence of the scattering matrix $\mathbf{\Phi}$ between consecutive iterations, measured by $\| \mathbf{\Phi} - \mathbf{\Phi}^{\text{prev}} \|_\infty$.}
    \label{fig:convergence_pdd}
\end{figure}
% We analyze the convergence of the proposed Algorithm \ref{alg:alg1} and demonstrate that it converges to at least a local optimal solution. 
The convergence of the designed algorithms are analyzed and we show that the value of the objective function \eqref{eq:op2} converges to at least a local optimal solution. 
Let $v^t$ denote the objective function value \eqref{eq:op2} at the $t^\mathrm{th}$ iteration. In Algorithm \ref{alg:alg1}, each block optimization problem is solved iteratively. A special block involves the PDD method for updating the scattering matrix $\mathbf{\Phi}_g$ for each group $g \in \mathcal{G}$, as formulated in the sub-optimization $\mathcal{P}8$. In the inner loop of the PDD method, the sub-problem w.r.t. $\mathbf{\Phi}_g$, \ie $\mathcal{P}9$, is convex and has a unique optimal solution. Similarly, the sub-problem w.r.t. $\mathbf{\Psi}_g$, \ie $\mathcal{P}10$, is also convex and has a unique optimal solution. Consequently, the value of the objective function \eqref{eq:op7} in $\mathcal{P}8$ is monotonically non-increasing. Since the objective function \eqref{eq:op7} is bounded below, the inner loop is guaranteed to converge to a local optimal solution. The outer loop of the PDD method converges to a KKT point because the augmented Lagrangian progressively enforces constraint satisfaction while updating the dual variables \cite{shi2020penalty}. Therefore, the solution for $\mathbf{\Phi}$ is guaranteed to be a local optimal solution. When the other variables are fixed, each block optimization problem for $\mathbf{\iota}$, $\mathbf{\tau}$, $\mathbf{P}$, and $\mathbf{\Phi}$ is convex and has a unique optimal solution. {Consequently, the value of the objective function \eqref{eq:op2} is monotonically non-decreasing after each iteration, \ie $v^t \geq v^{t-1}$. Since the objective function \eqref{eq:op2} is bounded above, it converges to a local optimum.}

We evaluate the convergence performance of the proposed Algorithm \ref{alg:alg1} and the PDD method in Algorithm \ref{alg:alg2} through simulations. We consider $K=3$ FD UEs, set the RIS element count to $M=32$, and include direct links to demonstrate the general applicability. The convergence results are provided in \fig{fig:convergence} (a) for fully-connected NR-BD-RIS, R-BD-RIS, and D-RIS. It is observed that Algorithm \ref{alg:alg1} works for all 3 types of RIS, and converges within $10$ iterations. The fully-connected NR-BD-RIS achieving the largest values. Additionally, \fig{fig:convergence} (b) presents the convergence performance for group-connected NR-BD-RIS. The values of objective function decrease as the group size $M_g$ becomes smaller.
The convergence of the PDD method, used to optimize $\mathbf{\Phi}$, is also analyzed.  In \fig{fig:convergence_pdd} (a), the difference between $\mathbf{\Phi}$ and its replica $\mathbf{\Psi}$ in the outer loop is shown. Convergence is achieved within $80$ iterations, with the equality constraint \eqref{eq:op70c2} being well satisfied, as indicated by $\| \mathbf{\Phi} - \mathbf{\Psi} \|_\infty < 10^{-6}$. \fig{fig:convergence_pdd} (b) presents the difference in $\mathbf{\Phi}$ between consecutive iterations, which decreases to approximately $10^{-3}$. To conclude, the PDD algorithm can converge faster if a smaller penalty parameter $c$ is adopted. Additionally, an appropriate range for the penalty parameter is $c \in (0.7, 0.95)$.

% \fig{fig:convergence} (a) illustrates the value of $v^t$ versus the number of iterations, confirming the convergence of the proposed algorithm. \fig{fig:convergence_pdd} shows the convergence of the PDD method, where the stop critera is reached. 

\subsection{Complexity Analysis}
Within the BCD framework, each block is updated iteratively. The computational complexity for updating $\boldsymbol{\iota}$ and $\boldsymbol{\tau}$ is $\mathcal{O}(K^2 M^2)$ per iteration. Updating the precoder $\mathbf{P}$ involves matrix inversion and bisection search, resulting in a complexity of $\mathcal{O}(K(M^2 + I_\mathsf{b} N^3))$, where $I_\mathsf{b}$ denotes the number of bisection search iterations. Similarly, updating the combiner $\mathbf{W}$ needs $\mathcal{O}(K N^3)$ operations because of the matrix inversion. PDD method to otpimize the scattering matrix $\mathbf{\Phi}$ is the most computationally intensive step.
Specifically, the computation of \eqref{eq:phi} involves the Kronecker product, contributing a complexity of $\mathcal{O}(M_g^2 M^4)$, while the SVD operation in \eqref{eq:phi_copy} adds $\mathcal{O}(M_g^3)$. As a result, the computational complexity of Algorithm \ref{alg:alg2} is $\mathcal{O}(G I_\mathsf{outer} I_\mathsf{inner} M_g^2 M^4)$, where $I_\mathsf{outer}$ is the outer iteration number, and $I_\mathsf{inner}$ is inner iteration number in the PDD method.
Combining these results, the total computational complexity of the iterative optimization algorithm is $\mathcal{O}(I_\mathsf{BCD} G I_\mathsf{outer} I_\mathsf{inner} M_g^2 M^4)$, where $I_\mathsf{BCD}$ denotes BCD iteration number needed for convergence. 

\begin{remark}
To improve scalability for larger RIS size and FD UE deployment, future work can build on the projected gradient descent method \cite{zhou2025joint}. Specifically, the scattering matrix $\boldsymbol{\Phi}$ with reciprocal and unitary constraints \eqref{eq:op1c4} and \eqref{eq:op1c3} lies in the closed set
$
\boldsymbol{\Phi} \in \mathcal{S}_{\boldsymbol{\Phi}}:=\left\{\boldsymbol{\Phi} \mid \boldsymbol{\Phi}=\boldsymbol{\Phi}^\top, \quad \boldsymbol{\Phi}^H \boldsymbol{\Phi}=\mathbf{I}\right\}.
$
The projection of a matrix $\mathbf{X}$ onto $\mathcal{S}_{\boldsymbol{\Phi}}$ has a closed form \cite{zhou2025joint}
$
\operatorname{Proj}_{\mathcal{S}_{\Theta}}(\mathbf{X})=\operatorname{uni}(\operatorname{sym}(\mathbf{X})),
$
where $\operatorname{sym}(\mathbf{X}):=\frac{1}{2}\left(\mathbf{X}+\mathbf{X}^T\right)=\arg \min _{\mathbf{Y}=\mathbf{Y}^T}\|\mathbf{Y}-\mathbf{X}\|_F^2$, $\operatorname{uni}(\mathbf{X}):=\mathbf{U V}^H=\arg \min _{\mathbf{Y Y}^H=\mathbf{I}}\|\mathbf{Y}-\mathbf{X}\|_F^2$, and $\mathbf{X}=\mathbf{U D V}^H$ is the SVD of $\mathbf{X}$.
The PGD update then applies this projection at each iteration, turning the double-loop PDD method into a single-loop PGD method with per-iteration complexity $\mathcal{O}(M^3)$. In addition, the partially proximal alternating direction method of Multipliers (ADMM) method \cite{wu2024optimization} can also improve scalability with per-iteration complexity $\mathcal{O}(K^3 M^3)$.
% Subsequently, the PGD update is utilized. This transform make the double-loop PDD method to a single-loop PGD method, which has per-iteration complexity of $\mathcal{O}\left(M^{3}\right)$. In addition, the partially proximal ADMM method \cite{wu2024optimization} can also improve the scalability with per-iteration complexity of $\mathcal{O}\left(K^{3} M^{3}\right)$.
\end{remark}
\color{black}

% We briefly analyze the computational complexity of the proposed algorithm in this section. Within the BCD framework, we iteratively update the blocks.  In each iteration, updating $\boldsymbol{\iota}_{\mathsf{d}}$, and $\boldsymbol{\tau}_{\mathsf{d}}$ require $\mathcal{O}(K^2 M^2)$ operations, and $\boldsymbol{\iota}_{\mathsf{u}}$, and $\boldsymbol{\tau}_{\mathsf{u}}$ require $\mathcal{O}(I^2 M^2)$ operations. The step for updating precoder $\mathbf{P}$ has the complexity $\mathcal{O}(K(M^2+I_p N^3))$ due to the matrix inversion and bisection search, where $I_p$ is the iteration number of bisection search. The step for updating combiner $\mathbf{W}$ requires $\mathcal{O}(I N^3)$ operations due to the matrix inversion. The most computationally intensive step in BCD is updating the scattering matrix $\mathbf{\Phi}$ using the PDD based Algorithm \ref{alg:alg2}. The calculations of \eqref{eq:phi} and \eqref{eq:phi_copy} require $\mathcal{O}(M_g^2 M^4)$, primarily due to the Kronecker product and $\mathcal{O}(M_g^3)$ operations due to SVD, respectively. Therefore, the computational complexity of Algorithm \ref{alg:alg2} is $\mathcal{O}(G I_1 I_2 (M_g^2 M^4))$, where $G$ is the total number of groups, $I_1$ is the number of PDD outer iteration, and $I_2$ is the number of PDD inner iterations in the PDD. Considering the BCD framework in Algorithm \ref{alg:alg1}, the total computational complexity of the entire optimization algorithm is $\mathcal{O}(I_3 G I_1 I_2 (M_g^2 M^4))$, where $I_3$ is the number of BCD iteration until convergence.

\section{Numerical evaluation}
\label{sec:simu}
{We use numerical results to explain the one-way secure transmission and eavesdropping prevention in this FD wireless circulator systems.} We also show the advantages of the NR-BD-RIS over the R-BD-RIS and D-RIS, considering both direct and reflected links. We investigate the impact of the number of RIS elements and the group size of the RIS on the sum-rate performance. Additionally, we compare the sum-rate performance of the NR-BD-RIS with the R-BD-RIS and D-RIS under varying relative locations of the FD UEs. {To provide further insights of the secure communications, we show the sum secrecy rate in the presence of both internal and external EVEs, illustrate the impinging and reflected beampatterns for the three types of RIS and discuss reasons behind the superior performance of the NR-BD-RIS and how eavesdropping is prevented.} We also evaluate weighed sum-rate region by varying the weights $\alpha_i, \forall i \in \mathcal{K}$. The  rate regions for the three UEs highlight the trade-offs in performance among them. Finally, we investigate the sum-rate performance of the NR-BD-RIS with varying numbers of TX and RX antennas at each FD UE and the number of FD UEs.

The proposed algorithm is applicable to scenarios both with and without structural scattering. Note that the presented algorithm is designed for the case with structural scattering. In fact, the case without structural scattering is simpler in terms of optimization design. To evaluate the performance without structural scattering, the algorithm can be modified by removing the $-\mathbf{I}$ term after $\mathbf{\Phi}$ in all channel expressions.
In the simulations, direct links are considered in the sum-rate performance evaluation with respect to the number of RIS elements. {In addition, we have ensured that the optimization algorithms for the three types of RIS converge for a fair comparison.} These results confirm that the NR-BD-RIS provides performance gains over R-BD-RIS and D-RIS, even in the presence of direct links. In subsequent simulations, to better emphasize the benefits of the NR-BD-RIS, we assume that all direct links are blocked.
% \vspace{-10pt}
\subsection{Simulation Environment}
\begin{figure}[t]
    \centering
    \includegraphics[width = 0.4\textwidth]{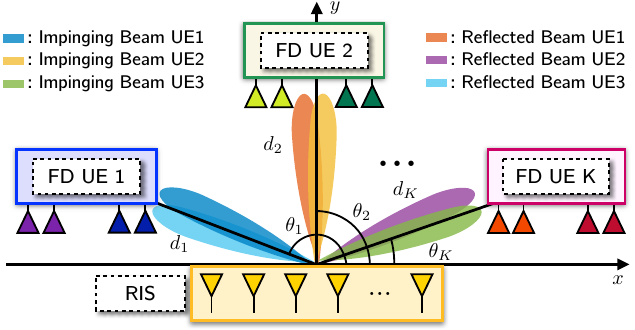}
    \centering
    \caption{2D coordinate system for the FD wireless circulator system enabled by RIS. The optimal impinging and reflected beampatterns are depicted.}
    \label{fig:simu_con}
\end{figure}
The 2D coordinate system of the FD wireless circulator with RIS is shown in \fig{fig:simu_con}. Each FD UE is equipped with $N$ TX and $N$ RX antennas. The channel models between the RIS and FD UEs include both large-scale and small-scale fading \cite{wu2019intelligent}. The large-scale fading is modeled using a distance-dependent path loss $\mathsf{P L}_i=\zeta_0\left(d_i / d_0\right)^{-\varepsilon_i}, \forall i \in \mathcal{K}$, where $\zeta_0$ is the attenuation factor at a reference distance $d_0 = 1$ m, $d_i$ is the distance between the RIS and FD UEs, and $\varepsilon_i$ is the path loss exponent for each UE. The small-scale fading is modeled using Rician fading, where the Rician factor $\kappa_i, \forall i \in \mathcal{K}$, represents the power ratio between the line-of-sight (LoS) component and the non-LoS component.

We set $\zeta_0=-30$ dB, with a path loss exponent of $\varepsilon_i = 2.2, \forall i \in \mathcal{K}$ for RIS-to-FD UE channels. The direct links are assumed to be weak, with a path loss exponent of $\varepsilon_i = 3.3$ \cite{zhou2023optimizing}. {Note that because the direct links are weak, this case yields lower sum-rate performance compared with the case using an RIS.} The distances between the RIS and the FD UEs are set to $d_1 = d_2 = d_3 = 35$ m. The Rician factor is set to $\kappa_i = 5$ to account for the LoS component. The transmit power of each FD UE is $P_i = 20$ dBm, $\forall i \in \mathcal{K}$, and the noise power at the FD UEs is $\sigma^2 = -80$ dBm. {Since SI is not the primary focus of this study, it is assumed to be well mitigated to the noise level. This is achieved by using cancellation techniques in the propagation domain \cite{everett2016softnull}, the analog domain \cite{debaillie2014analog}, and the digital domain \cite{liu2024full}.} {We assume perfect CSI at each FD UE in the simulation.}

\subsection{Sum-rates over RIS elements}
% \begin{figure}[t]
%     \centering
%     \includegraphics[width = 0.45\textwidth]{figure/ris.pdf}
%     \centering
%     \caption{The sum-rates versus number of RIS element of the three types of RIS without structural scattering with 3 users location at $30^\circ$, $90^\circ$, and $150^\circ$, Tx and Rx transmit antennas numbers $N = 1$.}
%     \label{fig:ris}
% \end{figure}
\begin{figure}[t]
    \centering
    \includegraphics[width = 0.48\textwidth]{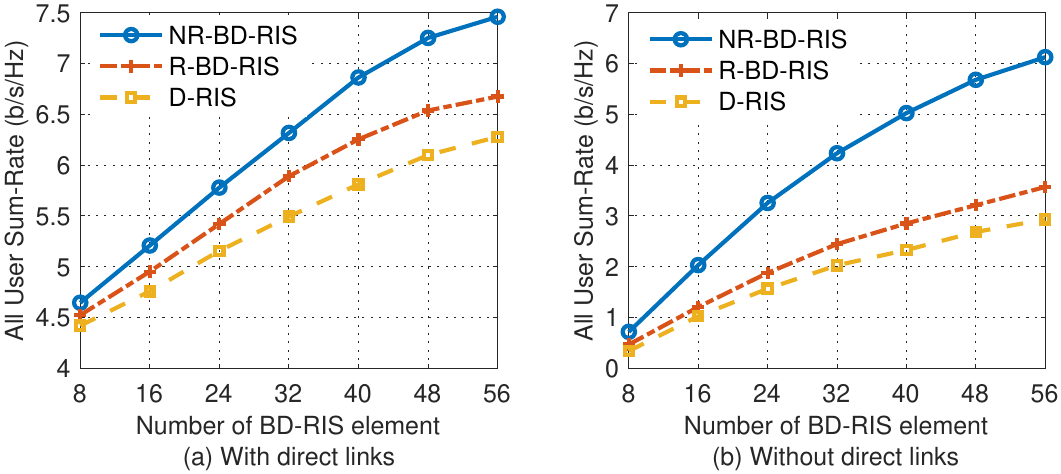}
    \centering
    \caption{Sum-rates versus the number of RIS element for the three RIS types: (a) with direct links and (b) without direct links. The NR-BD-RIS and R-BD-RIS are with fully-connected architecture. The 3 FD UEs are positioned at $30^\circ$, $90^\circ$, and $150^\circ$. Structural scattering is included, and the numbers of Tx and Rx antennas are $N = 1$.}
    \label{fig:ris}
\end{figure}
We analyze how number of RIS element affects the sum-rate performance. The 3 FD UEs are positioned at $30^\circ$, $90^\circ$, and $150^\circ$, respectively, with each equipped with $N = 1$ TX and RX antennas. \fig{fig:ris} (a) and (b) illustrate the sum-rates versus the number of RIS element, considering scenarios with and without direct links, respectively. The results show that the NR-BD-RIS consistently outperforms the R-BD-RIS and D-RIS. When number of RIS element increases, the sum-rate performance of the NR-BD-RIS improves, and its advantage over the other two RIS types becomes more obvious. This is attributed to the NR-BD-RIS's ability to break channel reciprocity and support multiple directions for FD UEs. Comparing the scenarios with and without direct links, the NR-BD-RIS maintains its superior performance. However, the performance gap between the NR-BD-RIS and the other two RIS types is smaller with direct links in \fig{fig:ris} (a), compared to the case without direct links, as shown in \fig{fig:ris} (b). This is because direct links provide additional communication paths for the FD UEs, reducing performance gain.

\subsection{Sum-rates}
To highlight the advantages of the NR-BD-RIS, we assume that the direct links are blocked, and the FD UEs connect exclusively via the RIS. This section presents the  sum-rates to showcase the superior performance of the NR-BD-RIS compared to the R-BD-RIS and D-RIS in the wireless circulator scenario.

\subsubsection{Sum-rates versus One Moving FD UE}
We assign equal weights to the three FD UEs, $\alpha_1 = \alpha_2 = \alpha_3 = 1/3$, and set the number of RIS element to $M=16$. Two scenarios are considered: in Case 1, FD UE 1 and FD UE 2 are fixed at $30^\circ$ and $90^\circ$, respectively, while in Case 2, their locations are set to $60^\circ$ and $105^\circ$. The location of FD UE 3 is then varied from $0^\circ$ to $180^\circ$ in increments of $1^\circ$. 

\begin{figure}[t]
    \centering
    \includegraphics[width = 0.48\textwidth]{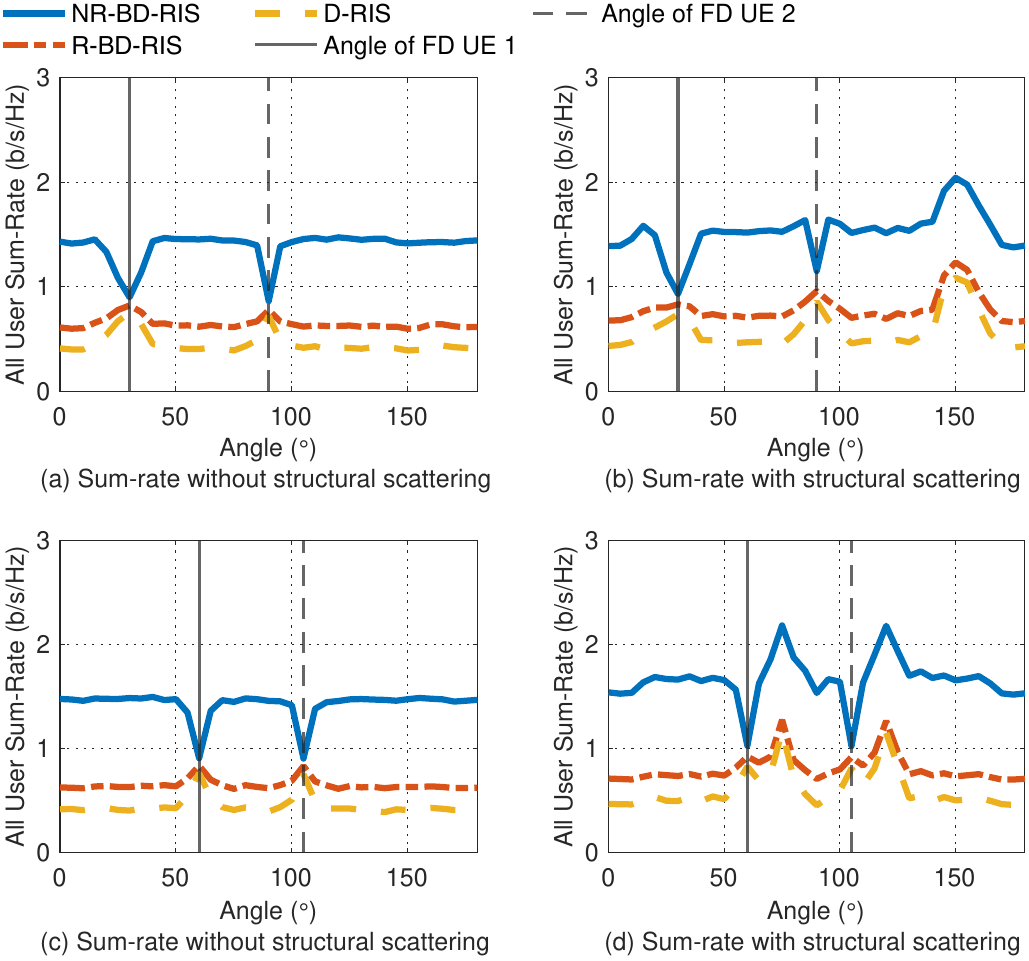}
    \centering
    \caption{Sum-rates for NR-BD-RIS, R-BD-RIS, and D-RIS with $M=16$ RIS elements and $N=1$ Tx/Rx antennas. (a), (b): FD UE 1 and UE 2 at $30^\circ$, $90^\circ$ without/with structural scattering. (c), (d): FD UE 1 and UE 2 at $60^\circ$, $105^\circ$ without/with structural scattering.}
    \label{fig:angle}
\end{figure}

In \fig{fig:angle} (a), the NR-BD-RIS consistently outperforms the R-BD-RIS and D-RIS in most cases. Specifically, the NR-BD-RIS has the same level of sum-rate performance as the other two RIS types when FD UE 3 is aligned with either FD UE 1 ($30^\circ$) or FD UE 2 ($90^\circ$). This is because R-BD-RIS and D-RIS can support one impinging direction and one reflected direction. However, when FD UE 3 is not aligned with FD UE 1 or FD UE 2, the NR-BD-RIS demonstrates a clear advantage, as it can support multiple directions, resulting in a higher  sum-rate. In \fig{fig:angle} (b), structural scattering is included, and the NR-BD-RIS still achieves the best sum-rate performance. 

Additionally, an enhancement is observed at the supplementary angle, \ie $150^\circ$, corresponding to FD UE 1's location ($30^\circ$). This specular reflection condition can be explained as follows. Considering the received power at the $k^\mathrm{th}$ UE, where the TX and RX antennas are set to $N = 1$, the received power is determined by the channel strength:
$\eta_k = |\mathbf{h}_k^\top (\mathbf{\Phi}-\mathbf{I})  \mathbf{h}_{k-1} |^2$.
% \begin{equation}
%     \label{eq:power}
%     \eta_k = |\mathbf{h}_k^\top (\mathbf{\Phi}-\mathbf{I})  \mathbf{h}_{k-1} |^2.
% \end{equation}
Using the triangle inequality, the Cauchy-Schwarz inequality, and the unitary condition $\mathbf{\Phi}^H \mathbf{\Phi} = \mathbf{I}$, the upper bound for the channel strength is derived as \cite{shen_modeling_2022, liu2024non}:
\begin{equation}
    | \mathbf{h}_k^\top (\mathbf{\Phi} - \mathbf{I}) \mathbf{h}_{k-1} |^2 \leq 
    ( \| \mathbf{h}_k  \|_2 \| \mathbf{h}_{k-1} \|_2 + | \mathbf{h}_k^\top \mathbf{h}_{k-1} | )^2.
\label{eq:bound}
\end{equation}
Equality holds when
\begin{equation}
    \beta \frac{\mathbf{h}_k^* }{\| \mathbf{h}_k  \|_2} = \mathbf{\Phi} \frac{\mathbf{h}_{k-1}}{\| \mathbf{h}_{k-1} \|_2}.
    \label{eq:eq_condition}
\end{equation}
The structural scattering term in \eqref{eq:bound} is represented by $| \mathbf{h}_k^\top \mathbf{h}_{k-1} | ^2$. For LoS channels, the amplitude of the structural scattering term is 
\vspace{-2pt}
\begin{equation}
    | \mathbf{h}_k^\top \mathbf{h}_{k-1} | = 1+ | \sum_{n = 1}^{M-1} e^{\jmath n \pi (\cos \theta_\mathsf{k} + \cos \theta_\mathsf{k-1} )}  |,
    \label{eq:strucatual_sca}
\end{equation}
where $\theta_\mathsf{k}$ and $\theta_\mathsf{k-1}$ are the angles of the $k^\mathrm{th}$ and $(k-1)^\mathrm{th}$ UEs, respectively. The maximum value of the structural scattering term is achieved when $\theta_\mathsf{k}+\theta_\mathsf{k-1} = \pi$. Therefore, a sum-rate enhancement is observed at the supplementary angle of FD UE 1 and FD UE 2.

If the locations of FD UE 1 and FD UE 2 are changed to $60^\circ$ and $105^\circ$, respectively, the sum-rate performance is shown in \fig{fig:angle} (c) and (d). The NR-BD-RIS still shows the best sum-rate. When FD UE 3 is not aligned with FD UE 1 or FD UE 2, the NR-BD-RIS outperforms the other two RIS types. Structural scattering also enhances the sum-rate performance at the supplementary angle of FD UE 1 and FD UE 2.
\color{black}

\subsection{Secure Communication Performance}
The proposed secure FD wireless circulator is designed to enforce one-way transmission and prevent eavesdropping among UEs. {In this subsection, two cases with internal EVEs and external EVEs are considered. In the first case, we analyze the received power of the eavesdropping signal at the $k^\mathrm{th}$ UE when it tries to intercept other UEs (\ie $\sum_{i \in \mathcal{K},, i \neq k,, k-1}$). For example, UE 1 attempts to eavesdrop UE 2 in a three UE case. A low received power shows that the NR-BD-RIS suppresses eavesdropping inside the wireless circulator and ensures secure communication. Next, we evaluate the secrecy sum rates of both internal and external EVEs to show the benefit of the NR-BD-RIS.} Additionally, we examine the impinging and reflected beampatterns of each FD UE to understand how one-way secure transmission and eavesdropping prevention are achieved. Ideally, the impinging beam of each FD UE should probe its own direction, while the reflected beam should probe the direction of the next FD UE, as illustrated in \fig{fig:simu_con}.

\subsubsection{{Received Power of Eavesdropping Signal}}
\begin{figure}[t]
    \centering
    \includegraphics[width = 0.4\textwidth]{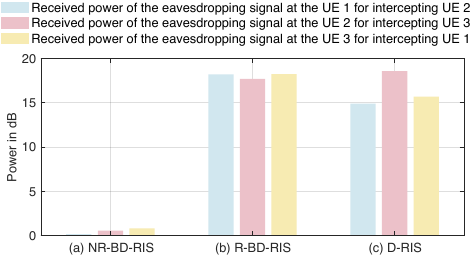}
    \centering
    \caption{{Received power of the eavesdropping signal at the $k^\mathrm{th}$ UE when it tries to intercept other $i^\mathrm{th}$ UEs (\ie $\sum_{i \in \mathcal{K}, i \neq k, k-1}$). For example, the blue bar means the received power of the eavesdropping signal for intercepting UE 2 at the UE 1.} The power is calculated in dB over noise power for three types of RISs with $M=32$ and $P_t = 40$ dBm. The 3 FD UEs with single antenna are positioned at $30^\circ$, $75^\circ$, and $120^\circ$, respectively.}
    \label{fig:muinter}
\end{figure}
{
We consider the EVEs are inside the wireless circulator. Each bar illustrates the received power of the eavesdropping signal at the $k^\mathrm{th}$ UE when it tries to intercept all other UEs (\ie ${i \in \mathcal{K}, i \neq k, k-1}$). This is also the MU interference from non-intended UEs from the perspective of interference. Note that $k^\mathrm{th}$ UE cannot eavesdrop itself (\ie $k^\mathrm{th}$) and the previous one (\ie $(k-1)^\mathrm{th}$), because $k^\mathrm{th}$ FD UE intends to receive signal from $(k-1)^\mathrm{th}$ FD UE.
For example, the blue bar means the received power of the eavesdropping signal for intercepting UE 2 at the UE 1.
As shown in \fig{fig:muinter}, the NR-BD-RIS achieves the lowest received power, effectively suppressing eavesdropping signals for other FD UEs.} This suppression ensures that no information leakage occurs, even if an FD UE attempts to eavesdrop on signals from others. The suppression capability can be further enhanced when multiple antennas are employed at the FD UEs, providing additional spatial degrees of freedom. {In contrast, the R-BD-RIS and D-RIS show higher received power of eavesdropping signals, indicating their limited capability to suppress eavesdropping.} \
{Note that the received power of D-RIS is sometimes smaller than R-BD-RIS, but the D-RIS's power of data signal is also smaller; this is because of the limited ability of D-RIS to manipulate waves. A more important metric is the rate shown in \fig{fig:ris}, \fig{fig:tx_risvssecrecy_internal} and \fig{fig:tx_risvssecrecy}, where the performance of D-RIS is worse than R-BD-RIS and NR-BD-RIS.}
The performance gain is due to the NR-BD-RIS's ability to break channel reciprocity, enabling greater flexibility in beam control compared to the R-BD-RIS and D-RIS.

\subsubsection{Secrecy Rate in the Presence of Internal Eavesdroppers}
\begin{figure}[t]
    \centering
    \includegraphics[width = 0.48\textwidth]{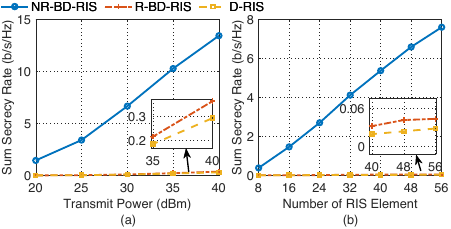}
    \centering
    \caption{{Sum secrecy rate in the presence of internal EVEs versus (a) the transmit power and (b) the number of RIS. The 3 FD UEs are positioned at $30^\circ$, $75^\circ$ and $120^\circ$. }}
    \label{fig:tx_risvssecrecy_internal}
\end{figure}
To measure the secure communication performance of the FD wireless circulator, we evaluate the sum secrecy rate defined as $\sum_{k \in \mathcal{K}} R_{\mathsf{s},k}^\mathsf{int}$ \eqref{eq:secrecyrate_interal} where the EVEs are internal. Specifically, the $k^\mathrm{th}$ FD UE are eavesdropped by $i^\mathrm{th}$ FD UE, where $i \in \mathcal{K}, i \neq k, k+1$. For example, UE 1 can be the internal eavesdropper for UE 2. The three FD UEs are positioned at $30^\circ$, $75^\circ$, and $120^\circ$. All direct links are assumed to be blocked. 

\fig{fig:tx_risvssecrecy_internal} (a) shows the sum secrecy rate versus the transmit power of each FD UE with internal EVEs, where the number of RIS elements is $M=16$. The NR-BD-RIS achieves the highest sum secrecy rate, followed by the conventional R-BD-RIS and D-RIS. In addition, the sum secrecy rates of R-BD-RIS and D-RIS are close to zero. This means that the UE rates are close to the rates of the eavesdroppers, so the information leakage is severe and there is no secure communication for R-BD-RIS and D-RIS. We attribute this to the fact that NR-BD-RIS breaks channel reciprocity, which enables uni-directional transmission and suppresses the eavesdropping signals inside the wireless circulator. For the same reason, as shown in \fig{fig:tx_risvssecrecy_internal} (b), NR-BD-RIS achieves the best sum secrecy rate and has large gaps compared with R-BD-RIS and D-RIS.

\subsubsection{Secrecy Rate in the Presence of External Eavesdroppers}
\begin{figure}[t]
    \centering
    \includegraphics[width = 0.48\textwidth]{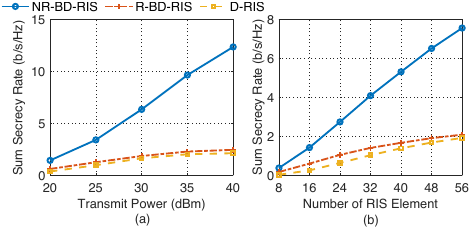}
    \centering
    \caption{{Sum secrecy rate in the presence of external EVEs versus (a) the transmit power and (b) the number of RIS. The 3 FD UEs are positioned at $30^\circ$, $75^\circ$ and $120^\circ$. The three EVEs with single antenna are located at $45^\circ$, $90^\circ$ and $135^\circ$.}}
    \label{fig:tx_risvssecrecy}
\end{figure}
The case is that there are external EVEs intercepting the wireless circulator. With the optimized variables from solving the sum-rate maximization problem $\mathcal{P}1$, we evaluate the sum secrecy rate in the presence of external EVEs. In this setup, three EVEs with single antenna are positioned at $45^\circ$, $90^\circ$, and $135^\circ$, respectively. The $k^{\mathrm{th}}$ EVE attempts to eavesdrop on the signal transmitted from the $k^{\mathrm{th}}$ FD UE. The sum secrecy rate of all FD UEs is defined as $\sum_{k \in \mathcal{K}} R_{\mathsf{s},k}^\mathsf{ext}$ \eqref{eq:secrecyrate} \cite{niu2021weighted}.

\fig{fig:tx_risvssecrecy} (a) illustrates the sum secrecy rate versus the transmit power of each FD UE with external EVEs, where the number of RIS elements is $M=16$. The NR-BD-RIS consistently achieves the highest sum secrecy rate across all transmit power levels, showing its effectiveness in improving secure communication. In comparison, the R-BD-RIS and D-RIS show lower sum secrecy rates due to their limited ability to break channel reciprocity and control beams. \fig{fig:tx_risvssecrecy} (b) shows the sum secrecy rate versus the number of RIS elements with external EVEs. The NR-BD-RIS outperforms the R-BD-RIS and D-RIS across all RIS sizes, showing its stronger capability in improving secure communication.

\color{black}

\subsubsection{Beampattern}
% \begin{figure}[t]
%     \centering
%     \includegraphics[width = 0.48\textwidth]{figure/bp48_0217.pdf}
%     \centering
%     \caption{Impinging and reflected beampatterns for RIS with $M=48$ elements: (i) without structural scattering and (ii) with structural scattering. Beampatterns for the three RIS types are shown as follows: (a), (d) NR-BD-RIS, (b), (e) R-BD-RIS, and (c), (f) D-RIS. The 3 FD UEs are located at $60^\circ$, $105^\circ$, and $150^\circ$, respectively. The Tx and Rx antenna numbers are $N = 1$.}
%     \label{fig:beam}
% \end{figure}
\begin{figure}[t]
    \centering
    \includegraphics[width = 0.5\textwidth]{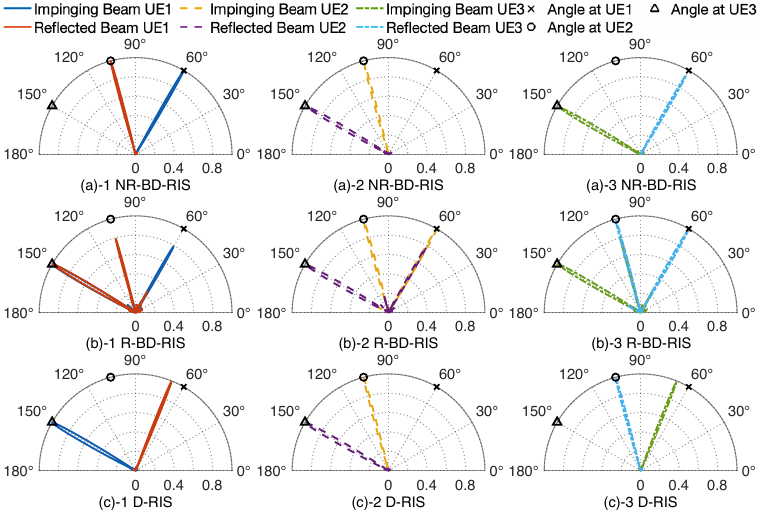}
    \centering
    \caption{Impinging and reflected beampatterns for RIS with $M=48$ elements without structural scattering. Beampatterns for the three RIS types are shown as follows: (a) NR-BD-RIS, (b) R-BD-RIS, and (c) D-RIS. The first, second, and third columns show the impinging and reflected beampatterns for UE 1, UE 2, and UE 3, respectively. The 3 FD UEs are located at $60^\circ$, $105^\circ$, and $150^\circ$, respectively. The Tx and Rx antenna numbers are $N = 1$.}
    \label{fig:beam}
\end{figure}
To better understand how PHY secrecy is achieved and why the NR-BD-RIS outperforms the other two RISs, we illustrate the impinging and reflected beampatterns for NR-BD-RIS, R-BD-RIS, and D-RIS. The beampatterns are shown in \fig{fig:beam}. The locations of the 3 FD UEs are $60^\circ$, $105^\circ$, and $150^\circ$, respectively. We set the number of RIS element to $M=48$. Specifically, the optimal directions for each beam are depicted in \fig{fig:simu_con}; the impinging beam of UE 1 should target the direction from UE 1 to the RIS, while the reflected beam of UE 1 should target the direction from the RIS to UE 2. The impinging and reflected beams for UE 2 and UE 3 follow the same principle.

The impinging and reflected beams for the $k^\mathrm{th}$ UE without structural scattering are defined as: 
$$
P_{k}^\mathsf{impinging}(\theta) =  |\mathbf{h}_{k+1}^\top \mathbf{\Phi} \mathbf{a}(\theta)|^2, \quad P_{k}^\mathsf{reflected}(\theta) =  | \mathbf{a}^\top(\theta) \mathbf{\Phi} \mathbf{h}_{k}|^2,
$$
where $\mathbf{a} = \frac{1}{\sqrt{N} } [1, e^{\jmath \pi \cos(\theta)}, \cdots, e^{\jmath \pi (N-1) \cos(\theta)}]^\top \in \mathbb{C}^{N \times 1}$ denotes the steering vector, and $\theta \in [0, 180^\circ]$. Normalization is applied to ensure the maximum value of the beampattern equals to $1$, specifically, \eg $P_{k}^\mathsf{impinging}(\theta)/P_{\mathsf{beam, max}}$, where $P_{\mathsf{beam, max}} \triangleq  \max \Big \{  \max\{ P_{k}^\mathsf{impinging} (\theta)  \}, \max\{ P_{k}^\mathsf{reflected}(\theta)  \} \Big \}, \forall k \in \mathcal{K}$.

The first column of \fig{fig:beam} shows the beampatterns for the impinging and reflected beams of UE 1. It can be observed that the impinging beam of UE 1 (\ie blue beam) correctly probes the direction of UE 1 at $60^\circ$, while the reflected beam of UE 1 (\ie red beam) correctly probes the direction of UE 2 at $105^\circ$. In comparison, the R-BD-RIS can probe the desired directions, but the beampattern power is lower than that of the NR-BD-RIS. {Part of the impinging beam probes UE 3, which is eavesdropping and can cause information leakage of UE 3. Also, part of the reflected beam probes the direction of UE 3, which can cause information leakage of itself. In addition, the impinging and reflected beams of the D-RIS probe unwanted directions, which can lead to information leakage and eavesdropping.}

The second and third columns of \fig{fig:beam} show the impinging and reflected beampatterns of UE 2 and UE 3, respectively. The NR-BD-RIS effectively targets the desired directions, while the R-BD-RIS and D-RIS can only target some of the desired directions. {In conclusion, the NR-BD-RIS can achieve one-way secure transmission due to its capability to support multiple beam directions. In contrast, the R-BD-RIS and D-RIS may cause information leakage and eavesdropping due to their limited control of two beam directions.} The results of the beampatterns also verify the unique benefit of the NR-BD-RIS in supporting multiple directions, which is reflected in better sum-rate performance compared to the R-BD-RIS and D-RIS, as shown in \fig{fig:ris} and \fig{fig:angle}.

\subsection{MU Sum-rate Region}
\begin{figure}[t]
    \centering
    \includegraphics[width = 0.48\textwidth]{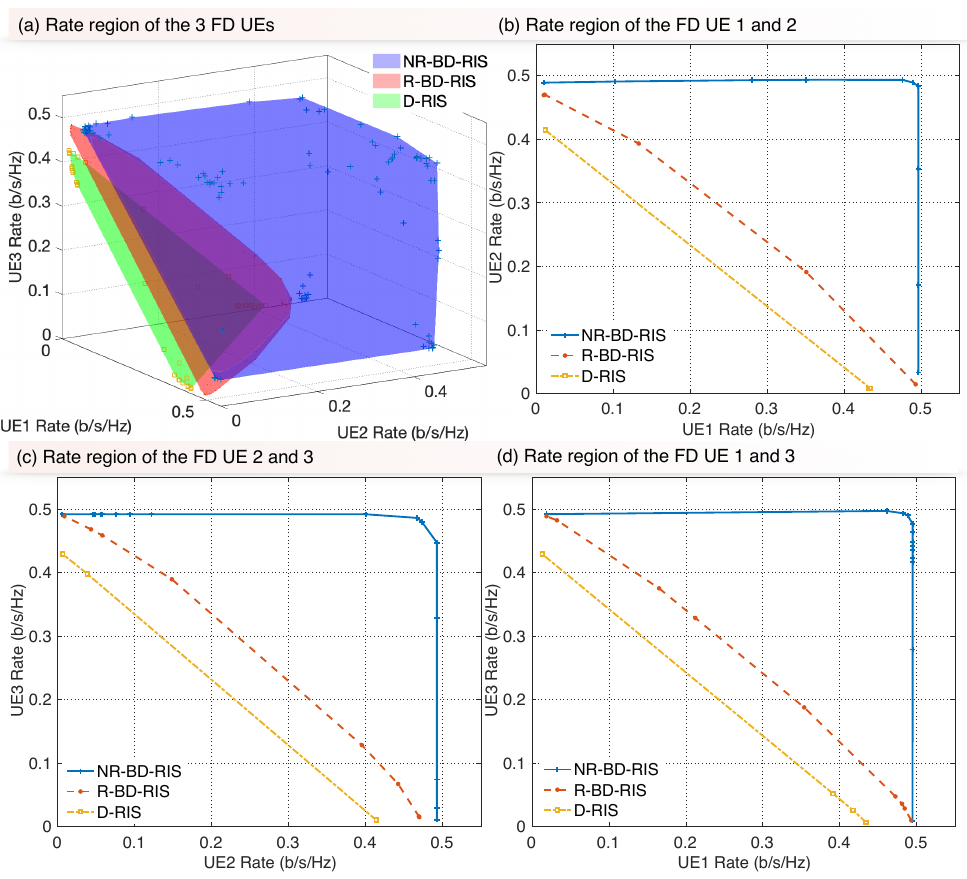}
    \centering
    \caption{Sum-rate regions for NR-BD-RIS, R-BD-RIS, and D-RIS with $M=16$ without structural scattering. The FD UE locations are $30^\circ$, $90^\circ$, and $150^\circ$, respectively, with $N = 1$ TX and RX antennas. (a) Sum-rate region for all 3 FD UEs, (b) Sum-rate region for UE 1 and UE 2, (c) Sum-rate region for UE 2 and UE 3, (d) Sum-rate region for UE 1 and UE 3.}
    \label{fig:alpha_3d}
\end{figure}
% \vspace{-5pt}
Given that the problem formulation $\mathcal{P}1$ focuses on the weighted sum-rate, we analyze the sum-rate region for the three UEs. The RIS is configured with $M=16$ elements, and TX and RX antennas at each UE are set to $N=1$. The UEs are positioned at $60^\circ$, $105^\circ$, and $150^\circ$, respectively. To emphasize the benefits of the NR-BD-RIS, the direct links are set to be blocked. The weights $\alpha_1, \alpha_2$, and $\alpha_3$ are varied between $0$ and $1$, ensuring $\sum \alpha_k = 1, \forall k \in \mathcal{K}$. The sum-rate region for all three UEs is illustrated in \fig{fig:alpha_3d} (a). Additionally, \fig{fig:alpha_3d} (b), (c), and (d) show the sum-rate regions for UE 1 and UE 2, UE 2 and UE 3, and UE 1 and UE 3, respectively. The results show that the NR-BD-RIS achieves the largest sum-rate region compared to the R-BD-RIS and D-RIS. This is attributed to the NR-BD-RIS's ability to break channel reciprocity and support multiple directions, whereas the R-BD-RIS and D-RIS are limited to one impinging and one reflected direction. {Furthermore, the rectangular shape of the NR-BD-RIS's sum-rate region indicates effective interference management, which enables suppression of signals from other UEs and thus prevents eavesdropping.}

\subsection{Sum-rates over Group-connected BD-RIS}
% \begin{figure}[t]
%     \centering
%     \includegraphics[width = 0.48\textwidth]{figure/gfix0623.pdf}
%     \centering
%     \caption{All-user sum-rates of NR-BD-RIS without structural scattering for varying group sizes $M_g$. In (a) and (b), the FD users are located at $30^\circ$, $90^\circ$, and $150^\circ$, respectively. In (c) and (d), the FD users are positioned at $60^\circ$, $105^\circ$, and $150^\circ$, respectively. The numbers of Tx and Rx antennas are set to $N = 1$.}
%     \label{fig:group}
% \end{figure}
\begin{figure}[t]
    \centering
    \includegraphics[width = 0.48\textwidth]{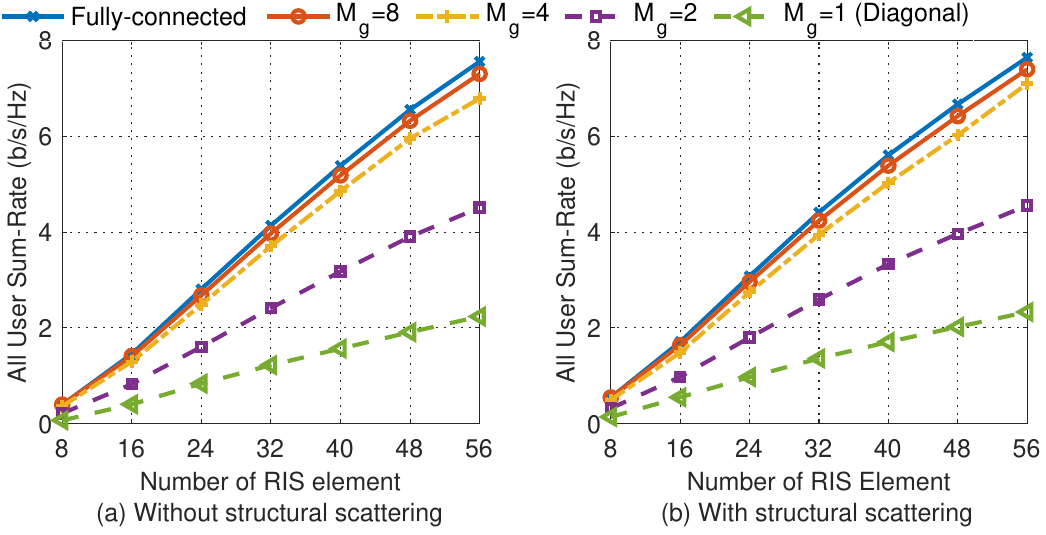}
    \centering
    \caption{Sum-rates of NR-BD-RIS for varying group sizes $M_g$. In (a) and (b), the FD UEs are positioned at $60^\circ$, $105^\circ$, and $150^\circ$, respectively. The numbers of Tx and Rx antennas are set to $N = 1$.}
    \label{fig:group}
\end{figure}
% We investigated the sum-rate performance of the NR-BD-RIS with different group sizes $M_g$. The locations of the 3 FD UEs are $30^\circ$, $90^\circ$, and $150^\circ$, respectively. The direct links are assumed to be blocked. The TX and RX antenna numbers are set to $N = 1$. The sum-rates versus the group size are shown in \fig{fig:group}. It can be observed that the sum-rate performance of the NR-BD-RIS increases with the group size. The upper bound is obtained by the fully-connected case, and the lower bound is achieved when the group size is $M_g = 1$, which corresponds to the D-RIS. We can also observe that with an increase in the number of RIS element, the gain among different group sizes increases.
We investigated the sum-rate performance of the NR-BD-RIS with different group sizes $M_g$. The locations of the 3 FD UEs are $60^\circ$, $105^\circ$, and $150^\circ$, respectively. The direct links are assumed to be blocked. The TX and RX antenna numbers are set to $N = 1$. The sum-rates versus the group size are shown in \fig{fig:group}. It can be observed that the sum-rate performance of the NR-BD-RIS increases with the group size. The upper bound is obtained by the fully-connected case, and the lower bound is achieved when the group size is $M_g = 1$, which corresponds to the D-RIS. We can also observe that with an increase in the number of RIS element, the gain among different group sizes increases.

\subsection{Sum-rates over TX and RX Antennas (MIMO Case)}
\begin{figure}[t]
    \centering
    \includegraphics[width = 0.48\textwidth]{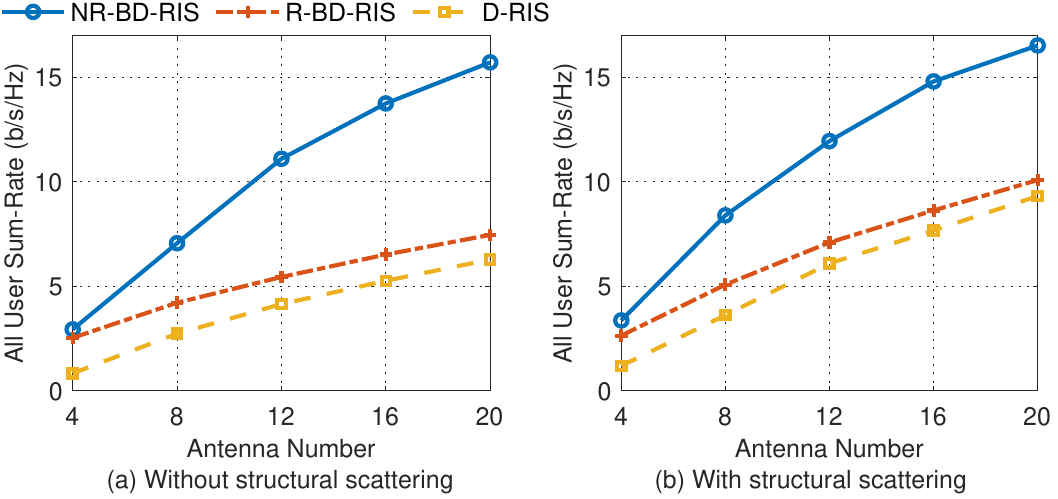}
    \centering
     \caption{The  sum-rates with MIMO setups (\ie $N_t = N_r = N$) versus different antenna numbers with $M=32$. The locations of the 3 FD UEs are $60^\circ$, $105^\circ$, and $150^\circ$, respectively. The results are shown for (a) without structural scattering and (b) with structural scattering.}
    \label{fig:antenna}
\end{figure}
As detailed in the system model in Section \ref{sec:sys} and the proposed algorithm in Section \ref{alg:alg1}, the transmit beamformer and receive combiner are jointly optimized, allowing the performance of MIMO setups to be observed. \fig{fig:antenna} illustrates the sum-rate performance with MIMO setups at each FD UE for the NR-BD-RIS, R-BD-RIS, and D-RIS as the number of TX and RX antennas varies. We set $N_t = N_r = N$. The 3 FD UEs are positioned at $60^\circ$, $105^\circ$, and $150^\circ$, respectively, with blocked direct links. The results show that the sum-rate performance of the NR-BD-RIS improves with an increasing number of antennas. Moreover, the performance gap between the NR-BD-RIS and the other two RIS types widens as the number of antennas increases. Additionally, comparing the case with structural scattering (\fig{fig:antenna} (b)) to the case without it (\fig{fig:antenna} (a)), the NR-BD-RIS achieves higher sum-rate performance when structural scattering is included. This is because the physics-compliant model, which incorporates structural scattering, leads to higher channel gains \cite{nerini2024physics}, thereby enhancing the sum-rate.

\subsection{Sum-rates over the Number of FD UEs}
\begin{figure}[t]
    \centering
    \includegraphics[width = 0.48\textwidth]{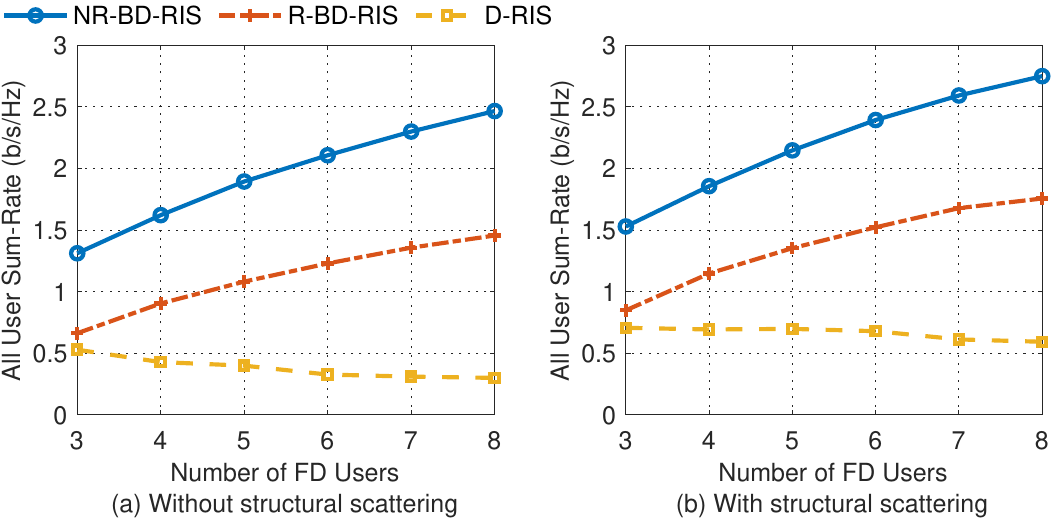}
    \centering
     \caption{The  sum-rates of NR-BD-RIS, R-BD-RIS, and D-RIS versus the number of FD UEs $K$ with $M=16$. The FD UEs are randomly located in the range of $(0^\circ, 180^\circ)$. The numbers of Tx and Rx antennas are set to $N = 1$. (a) Without structural scattering, (b) with structural scattering.}
    \label{fig:mu}
\end{figure}

We further analyze the sum-rate performance of the RISs as the number of FD UEs $K$ increases. The FD UEs are randomly located within the angular range of $(0^\circ, 180^\circ)$, and the RIS has $M=16$ elements. The results are shown in \fig{fig:mu}. Each FD UE is equipped with $N = 1$ TX and RX antennas. The results indicate that the NR-BD-RIS consistently outperforms the R-BD-RIS and D-RIS. Moreover, the sum-rate performance increases as the number of FD UEs increases. We attribute this to the NR-BD-RIS's capability to support multiple directions, enabling that more FD UEs can be supported. In contrast, the R-BD-RIS and D-RIS cannot fully support all directions, leading to worse performances. As the number of FD UEs increases, the sum-rate of the D-RIS decreases due to increased interference among the UEs. Additionally, when structural scattering is included (cf. \fig{fig:mu} (b)), the NR-BD-RIS achieves higher sum-rate performance compared to the case without structural scattering (cf. \fig{fig:mu} (a)). This enhancement is due to the increased channel gain provided by structural scattering \cite{nerini2024physics}.

\section{Conclusion}
\label{sec:con}
In this work, we have explored a secure FD wireless circulator system enabled by NR-BD-RIS, {facilitating one-way secure transmission and preventing eavesdropping among FD UEs.} The system model considers a physics-compliant channel model that incorporates structural scattering, loop interference, and SI. Structural scattering is essential for accurately modeling the RIS-assisted FD wireless circulator system. We have then formulated the weighted  sum-rate maximization problem. To solve this problem, we have proposed an iterative algorithm based on BCD and PDD. The numerical evaluations have shown the advantages of NR-BD-RIS over R-BD-RIS and D-RIS in terms of sum-rate performance due to its capability to break channel reciprocity. We attribute the benefit of the NR-BD-RIS to its ability to support multiple directions, while the R-BD-RIS and D-RIS are limited to one impinging and one reflected direction. The gain of NR-BD-RIS over R-BD-RIS and D-RIS increases with the number of RIS elements, group size, TX and RX antennas, and FD UEs due to more flexibility. This work has provided insights into the design of NR-BD-RIS for new applications in secure communication. {A detailed study on channel estimation and the impact of imperfect CSI on the proposed wireless circulator, as well as the maximization of the sum secrecy rate for enhanced secure communication are meaningful directions for future research.} {In addition, incorporating non-idealities due to hardware constraints into the system design is an important topic for future work.}

\bibliographystyle{IEEEtran_url}
\bibliography{IEEEabrv,references}

\end{document}